\documentclass[12pt]{article}
\usepackage{a4wide,epsfig,psfrag,amsmath,amssymb,cite,scalefnt}
\usepackage{color}
\usepackage{amsmath,comment,braket}

\graphicspath{ {figs/} }

\newcommand{\be}{\begin{equation}}
\newcommand{\ee}{\end{equation}}
\newcommand{\bea}{\begin{eqnarray}}
\newcommand{\eea}{\end{eqnarray}}
\newcommand{\ep}{\epsilon}

\newcommand{\mt}{m_t}
\newcommand{\mts}{m_t^2}
\newcommand{\nh}{n_h}
\newcommand{\nl}{n_l}
\newcommand{\ca}{C_A}
\newcommand{\cf}{C_F}

\newcommand{\cftwo}{C_F^2}
\newcommand{\cfca}{C_AC_F}
\newcommand{\catwo}{C_A^2}
\newcommand{\cfnl}{C_Fn_l}
\newcommand{\canl}{C_An_l}
\newcommand{\cfnh}{C_Fn_h}
\newcommand{\canh}{C_An_h}
\newcommand{\nltwo}{n_l^2}
\newcommand{\nlnh}{n_ln_h}

\newcommand{\Litwo}{\mbox{Li}_{2}}
\newcommand{\mHfour}{m_H^4}
\newcommand{\mHtwo}{m_H^2}

\newcommand{\LLst}{L_{st}}
\newcommand{\LLsu}{L_{su}}
\newcommand{\LLmttwos}{L_{m_t^2s}}
\newcommand{\LLmHtwot}{L_{m_H^2t}}
\newcommand{\LLmHtwou}{L_{m_H^2u}}

\allowdisplaybreaks

\newcommand{\gsim}{\;\rlap{\lower 3.5 pt \hbox{$\mathchar \sim$}} \raise 1pt
 \hbox {$>$}\;}
\newcommand{\lsim}{\;\rlap{\lower 3.5 pt \hbox{$\mathchar \sim$}} \raise 1pt
 \hbox {$<$}\;}

\begin{document}

\title{\vskip-3cm{\baselineskip14pt
    \begin{flushleft}
      \normalsize P3H-19-030\\
      \normalsize TTP19-028\\
    \end{flushleft}}
  \vskip1.5cm
  Three-loop form factors for Higgs boson pair production
  in the large top mass limit
}

\author{
  Joshua Davies and
  Matthias Steinhauser,
  \\[1mm]
  {\small\it Institut f{\"u}r Theoretische Teilchenphysik}\\
  {\small\it Karlsruhe Institute of Technology (KIT)}\\
  {\small\it Wolfgang-Gaede Stra\ss{}e 1, 76131 Karlsruhe, Germany}
}
  
\date{}

\maketitle

\thispagestyle{empty}

\begin{abstract}

  We consider the virtual corrections to Higgs boson pair production at
  next-to-next-to-leading order, in the large top quark mass limit.
  We compute five expansion terms for the box-type form factors and
  eight expansion terms for the triangle form factor, which serve as useful
  input for the construction of approximations. We present analytic results
  for the form factors in the soft-virtual approximation.

  From a technical point of view the calculation is quite challenging since
  huge intermediate expressions are produced. We describe our methods and
  optimizations to overcome these difficulties, which might be useful for other
  calculations.

%

\end{abstract}

\thispagestyle{empty}

\sloppy


\newpage


\section{\label{sec::intro}Introduction}

The simultaneous production of two Higgs bosons is a promising process to
obtain information about the self-coupling of the Higgs boson and thus the
structure of the scalar potential.  Although it is experimentally very
challenging it is expected that this process can be observed after the
high-luminosity upgrade of the LHC.

On the theoretical side there has been quite some effort to obtain precise
predictions for differential and total cross sections for Higgs boson pair
production. In analogy to single Higgs production the
numerically most important contribution is provided by gluon fusion, followed by
vector boson fusion, associated production with top quarks and the
Higgs-strahlung process (see, e.g., Ref.~\cite{Baglio:2012np}).

Exact leading order (LO) results for $gg\to HH$ have been available for more than
thirty years~\cite{Glover:1987nx,Plehn:1996wb}. Next-to-leading order (NLO)
corrections have been computed numerically much more recently,
and are available from two independent
groups~\cite{Borowka:2016ehy,Borowka:2016ypz,Baglio:2018lrj}.  Note that the
numerical evaluations are quite expensive. For this reason it is important to
have approximations at hand, which are valid in certain regions of the phase
space. Among them are large top quark mass
expansions~\cite{Dawson:1998py,Grigo:2014jma,Degrassi:2016vss} which are
available up to order $1/m_t^{12}$~\cite{Grigo:2014jma}.  Furthermore, in
Ref.~\cite{Bonciani:2018omm} an expansion around small transverse momentum has
been performed and results in the high-energy region are available
from~\cite{Davies:2018ood,Davies:2018qvx}. 
They have been combined in Ref.~\cite{Davies:2019dfy} with the exact
calculation from~\cite{Borowka:2016ehy,Borowka:2016ypz} to provide a precise
grid for the NLO virtual corrections~\cite{hhgrid}.
In Ref.~\cite{Maltoni:2014eza}
exact results for the real radiation contribution have been combined with the
effective-theory virtual corrections.  Interesting approximations for
$gg\to HH$ at NLO have been constructed in Ref.~\cite{Grober:2017uho} where
expansion terms from various regions have been combined with the help of a
conformal mapping and Pad\'e approximation. The same method has been been
applied in Ref.~\cite{Davies:2019nhm} (using the triangle form factor results
of this paper) to the Higgs-gluon form factor, an
important ingredient of single-Higgs boson production, in order
to reconstruct the full quark mass dependence.\footnote{Note that analytic
  results for the light fermion contribution to the three-loop Higgs-gluon form
  factor have been obtained in Ref.~\cite{Harlander:ggHnl}.}

At NNLO exact results are currently out of range, which makes it even more
important to obtain approximations, if possible from various kinematic
regions.  Within the effective theory, where the top quark mass is assumed to
be infinitely heavy, NNLO corrections have been computed in
Refs.~\cite{deFlorian:2013uza,deFlorian:2013jea,Grigo:2014jma}.
Power-suppressed terms have been obtained in Ref.~\cite{Grigo:2015dia}, where
the soft-virtual approximation was constructed.  Real corrections which
originate from three closed top quark loops have been computed in Ref.~\cite{Davies:2019xzc}. In
Ref.~\cite{Grazzini:2018bsd} approximate NNLO expressions are constructed on
the basis of the exact NLO results~\cite{Borowka:2016ypz} and further NNLO
building blocks which are also available for finite top quark mass. Other NNLO
contributions, such as the three-loop virtual corrections, are taken in the
infinite top quark mass limit. The results of this paper provide additional
$1/\mts$ corrections to the three-loop $gg\to HH$ amplitude which could improve
the approximations of Ref.~\cite{Grazzini:2018bsd}.

The resummation of threshold-enhanced logarithms to next-to-next-to-leading
logarithmic (NNLL) accuracy has been performed in
Refs.~\cite{Shao:2013bz,deFlorian:2015moa} and differential distributions
up to NNLO for various observables were computed in
Ref.~\cite{deFlorian:2016uhr} in the heavy-top limit. More recently,
finite top quark mass effects have also been included~\cite{deFlorian:2018tah}.

At N$^3$LO first results are available in the limit of
infinitely heavy top quarks. In Ref.~\cite{Banerjee:2018lfq} massless two-loop
box contributions have been computed and four-loop corrections to the
effective coupling of two Higgs bosons and two, three or four gluons became
available from~\cite{Spira:2016zna,Gerlach:2018hen}.

In this work we consider NNLO virtual corrections to $gg\to HH$ and compute
the three relevant form factors for a large top quark mass.  We evaluate five
expansion terms for the box-type form factors and eight expansion terms for the
triangle form factor, i.e., up to order $1/m_t^{8}$ and $1/m_t^{14}$, respectively.
The results for the two box-type form factors are new. The results for the triangle
form factor have been obtained in Ref.~\cite{Harlander:2009bw,Pak:2009bx} up to order
$1/m_t^{8}$, the higher-order expansion terms presented here are new.
In a previous
work~\cite{Grigo:2015dia} expansion terms up to $1/m_t^{4}$ were computed
for the (differential) cross section, but not for the form factors. Our
results constitute important input for the construction of approximations. For
example, it is possible to extend the consideration of
Ref.~\cite{Grober:2017uho} to NNLO.  Furthermore, as already mentioned above,
it might be possible to improve the approximations of
Ref.~\cite{Grazzini:2018bsd}.

The remainder of the paper is organized as follows: in the next section we
introduce our notation and define the form factors. We provide technical details
in Section~\ref{sec::calc}, and mention several optimizations which were crucial
to be able to perform the calculations. Ultraviolet
renormalization and infrared subtraction are discussed in
Section~\ref{sec::ren} and both analytical and numerical results are shown in
Section~\ref{sec::res}. We conclude in Section~\ref{sec::con}.


\section{\label{sec::setup}Setup}

The amplitude for the process $g(q_1)g(q_2)\to H(q_3)H(q_4)$
is conveniently decomposed into three form factors. In the following
we outline their precise definition. We start with the amplitude
which is given by
\begin{eqnarray}
  {\cal M}^{ab} &=& 
  \varepsilon_{1,\mu}\varepsilon_{2,\nu}
  {\cal M}^{\mu\nu,ab}
  \,\,=\,\,
  \varepsilon_{1,\mu}\varepsilon_{2,\nu}
  \delta^{ab}
  \left( {\cal M}_1 A_1^{\mu\nu} + {\cal M}_2 A_2^{\mu\nu} \right)
  \,,
\end{eqnarray}
where $a$ and $b$ are adjoint colour indices and the two Lorentz structures
are given by
\begin{eqnarray}
  A_1^{\mu\nu} &=& g^{\mu\nu} - {\frac{1}{q_{12}}q_1^\nu q_2^\mu
  }\,,\nonumber\\
  A_2^{\mu\nu} &=& g^{\mu\nu}
                   + \frac{1}{p_T^2 q_{12}}\left(
                   q_{33}    q_1^\nu q_2^\mu
                   - 2q_{23} q_1^\nu q_3^\mu
                   - 2q_{13} q_3^\nu q_2^\mu
                   + 2q_{12} q_3^\mu q_3^\nu \right)\,,
 \label{eq::A1A2}
\end{eqnarray}
with
\begin{eqnarray}
\label{eqn:qt2def}
  q_{ij} &=& q_i\cdot q_j\,,\qquad
  p_T^{\:2} \:\:\:=\:\:\: \frac{2q_{13}q_{23}}{q_{12}}-q_{33}
  \,.
\end{eqnarray}
${\cal M}_1$ and ${\cal M}_2$ can be projected from ${\cal M}^{\mu\nu}$ using
the projectors
\begin{eqnarray}
\label{eqn:FFprojectors}
  P_{1,\mu\nu} 
  &=&
      -\, q_{1,\nu} q_{2,\mu} \Bigg(
      \frac{1}{q_{12}} \frac{1-\epsilon}{2-4\epsilon}
      -\frac{q_{33}}{q_{12} p_T^{\:2}} \frac{\epsilon}{2-4\epsilon}
      \Bigg)
    +\, q_{1,\nu} q_{3,\mu}
	\Bigg(
		-\frac{2 q_{23}}{q_{12}p_T^2}\frac{\epsilon}{2-4\epsilon}
	\Bigg)
\nonumber\\&&
	+\, q_{2,\mu} q_{3,\nu}
	\Bigg(
		-\frac{2 q_{13}}{q_{12}p_T^2}\frac{\epsilon}{2-4\epsilon}
	\Bigg)
	-\, q_{3,\mu} q_{3,\nu}
	\Bigg(
		-\frac{2}{p_T^2}\frac{\epsilon}{2-4\epsilon}
	\Bigg)
	+\, g_{\mu\nu} q_{12}\Bigg(
	   	 \frac{1}{q_{12}}\frac{1}{2-4\epsilon}
	\Bigg)\,,
\nonumber\\[2mm]
 P_{2,\mu\nu} &=&
  	q_{1,\nu} q_{2,\mu}
  	\Bigg(
  		\frac{q_{33}}{q_{12}p_T^2} \frac{1-\epsilon}{2-4\epsilon}
  		-\frac{1}{q_{12}}\frac{\epsilon}{2-4\epsilon}
  	\Bigg)
    -\, q_{1,\nu} q_{3,\mu}
  	\Bigg(
	  	\frac{q_{23}}{q_{12}p_T^2} \frac{1-\epsilon}{1-2\epsilon}
  	\Bigg)
\nonumber\\&&
	-\, q_{2,\mu} q_{3,\nu}
  	\Bigg(
	  	\frac{q_{13}}{q_{12}p_T^2} \frac{1-\epsilon}{1-2\epsilon}
  	\Bigg)
	+\, q_{3,\mu} q_{3,\nu} 
  	\Bigg(
	  	\frac{1}{p_T^2} \frac{1-\epsilon}{1-2\epsilon}
  	\Bigg)
	+\, g_{\mu\nu} q_{12} \Bigg(
    	 \frac{1}{q_{12}}\frac{1}{2-4\epsilon}
	\Bigg)\,,
\end{eqnarray}
where $\epsilon = (4-d)/2$ is the standard dimensional regularization parameter.

The Feynman diagrams involving the triple-Higgs boson coupling only
contribute to $A_1^{\mu\nu}$, which is the only structure relevant for
single-Higgs production, therefore it is convenient to decompose
${\cal M}_1$ and ${\cal M}_2$ into ``triangle'' and ``box'' form factors
\begin{eqnarray}
  {\cal M}_1 &=& X_0 \, s \, \left(\frac{3 m_H^2}{s-m_H^2} F_{\rm tri} + F_{\rm box1}\right)
                 \,,\nonumber\\
  {\cal M}_2 &=& X_0 \, s \, F_{\rm box2}
                 \,,
                 \label{eq::calM}
\end{eqnarray}
with the prefactor
\begin{eqnarray}
  X_0 &=& \frac{G_F}{\sqrt{2}} \frac{\alpha_s(\mu)}{2\pi} T n_h \,,
          \label{eq::X0}
\end{eqnarray}
where $T=1/2$ and $n_h=1$ have been introduced for convenience.
We furthermore define the expansion of the form factors in $\alpha_s$ as
\begin{eqnarray}
  F_X &=& \sum_{i\ge0} \left(\frac{\alpha_s(\mu)}{\pi}\right)^i F^{(i)}_X
  \,,
  \label{eq::F}
\end{eqnarray}
with $X\in\{\mbox{tri},\mbox{box1},\mbox{box2}\}$.
Note that $F^{(i)}_X$ corresponds to the $(i+1)$-loop result.
In our final expressions the strong coupling constant is defined with five
active quark flavours, which is an appropriate choice since we consider
the top quark mass to be large.

In the course of the calculation it is convenient to introduce the
Mandelstam variables
\begin{eqnarray}
  {s}=(q_1+q_2)^2\,,\qquad {t}=(q_1+q_3)^2\,,\qquad {u}=(q_2+q_3)^2\,,
  \label{eq::stu}
\end{eqnarray}
with
\begin{eqnarray}
  q_1^2=q_2^2=0\,,\qquad  q_3^2=q_4^2=m_H^{2}\,,\qquad
  s+t+u=2m_H^{2}\,.
  \label{eq::q_i^2}
\end{eqnarray}
It is furthermore convenient to express the final result
in terms of the transverse momentum of one of the
Higgs bosons which is given in terms of the Mandelstam variables by
(equivalent to Eq.~(\ref{eqn:qt2def}))
\begin{eqnarray}
  p_T^2 &=& \frac{tu-m_H^4}{s}\,.
\end{eqnarray}


\section{\label{sec::calc}Calculation details}

\begin{figure}[t]
  \centering
  \begin{tabular}{cccc}
    \includegraphics[width=0.18\textwidth]{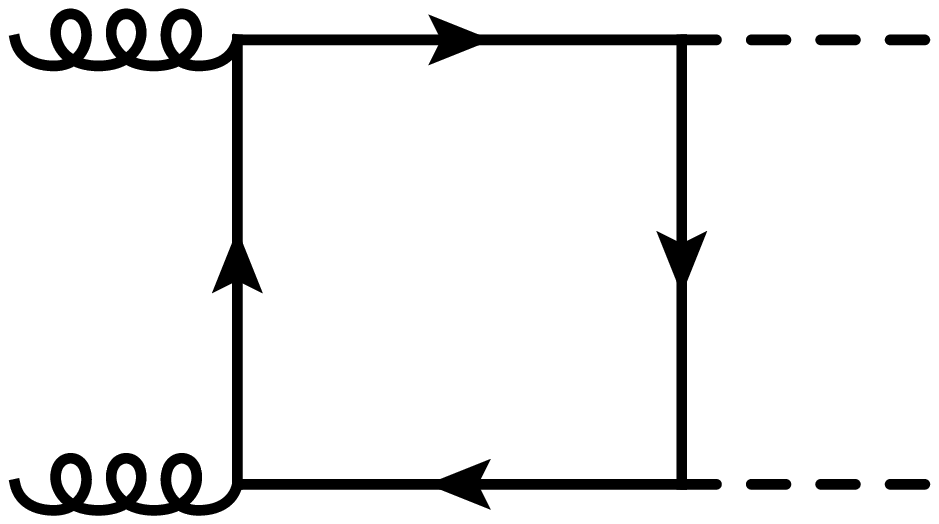} &
    \includegraphics[width=0.18\textwidth]{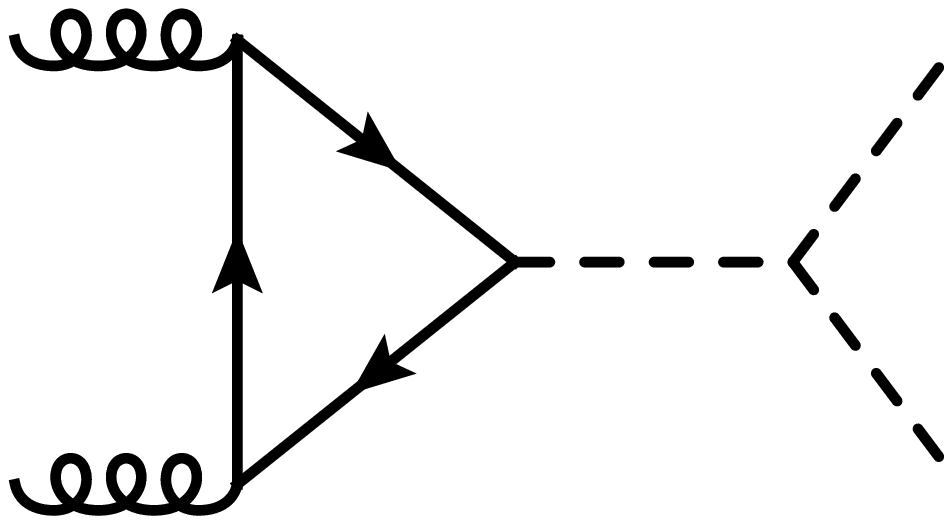} &
    \includegraphics[width=0.18\textwidth]{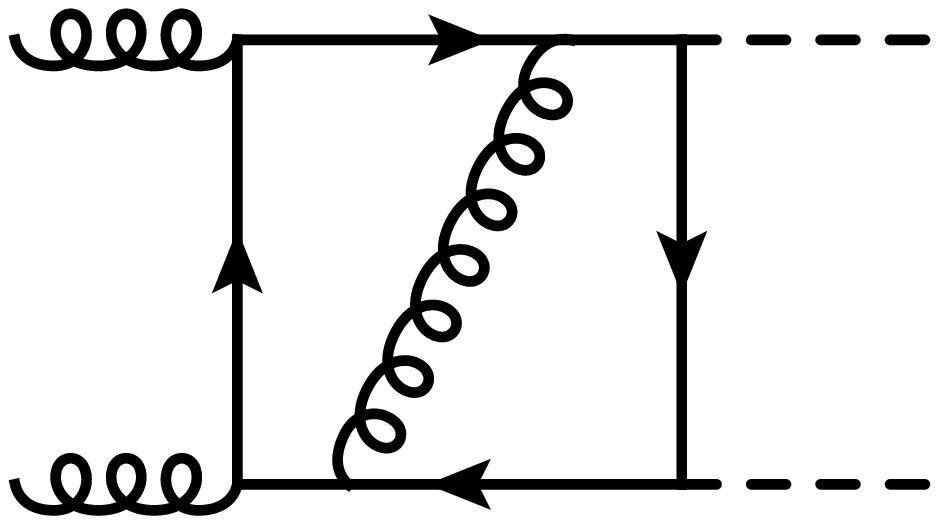}  &
    \includegraphics[width=0.18\textwidth]{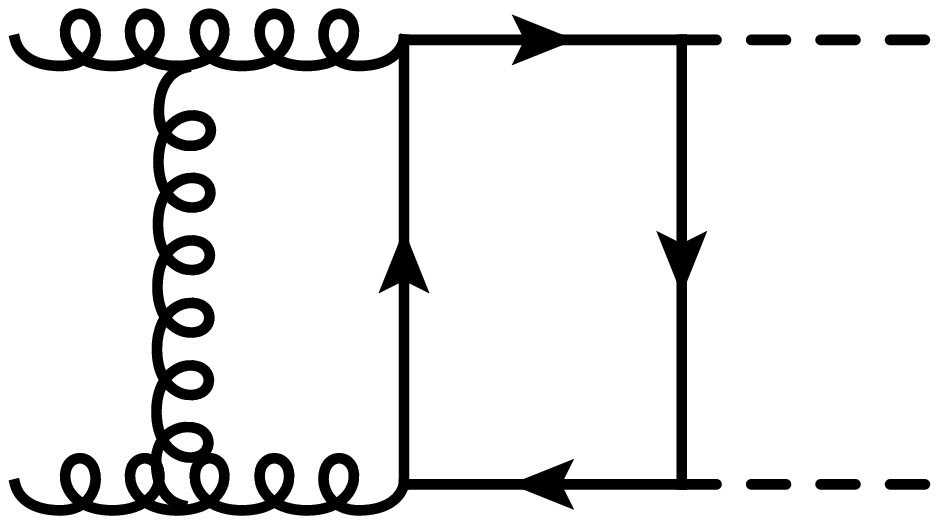}
    \\
    $Tn_h$ & $Tn_h$ & $C_FTn_h$ & $C_ATn_h$
    \\
    \includegraphics[width=0.15\textwidth]{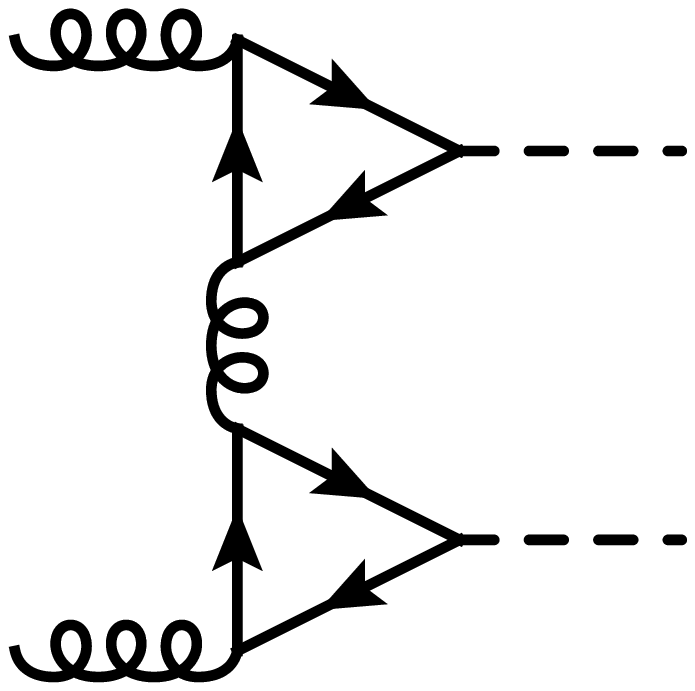}   &
    \raisebox{1.2em}{\includegraphics[width=0.18\textwidth]{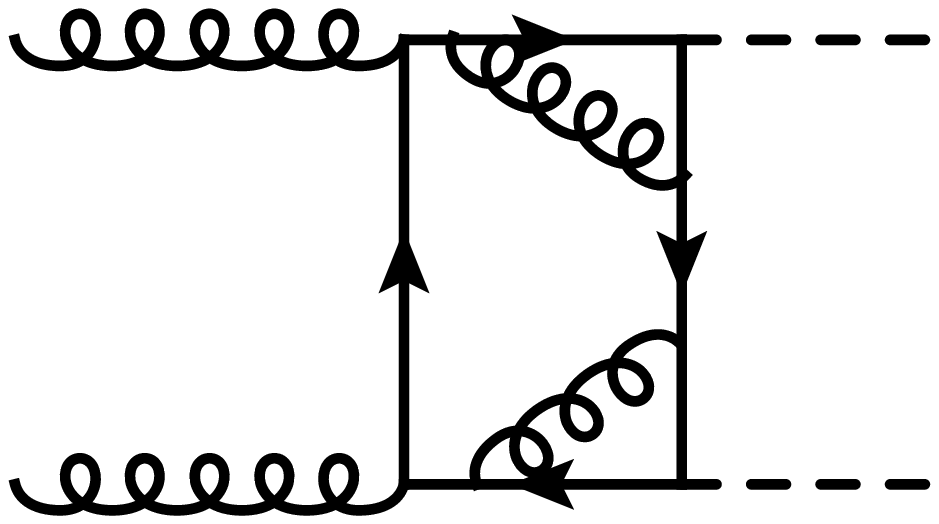}} &
    \raisebox{1.2em}{\includegraphics[width=0.18\textwidth]{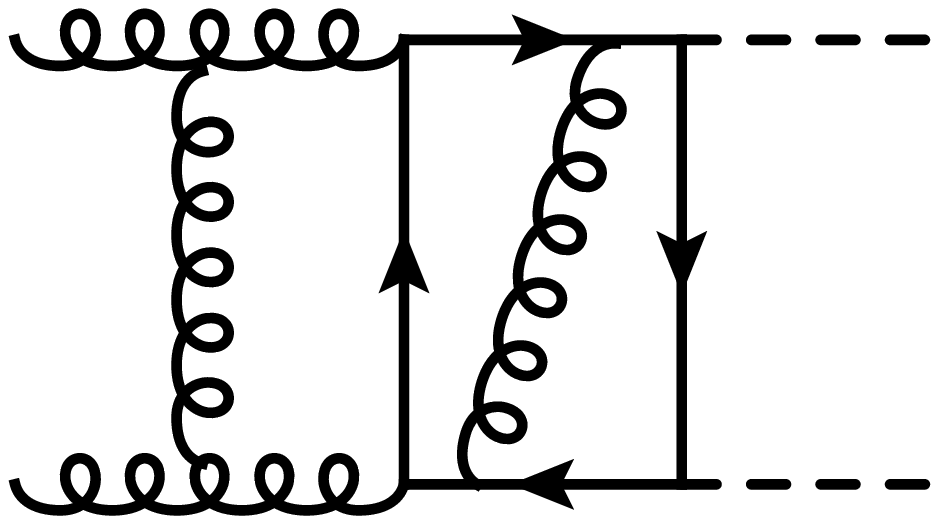}} &
    \raisebox{1.2em}{\includegraphics[width=0.18\textwidth]{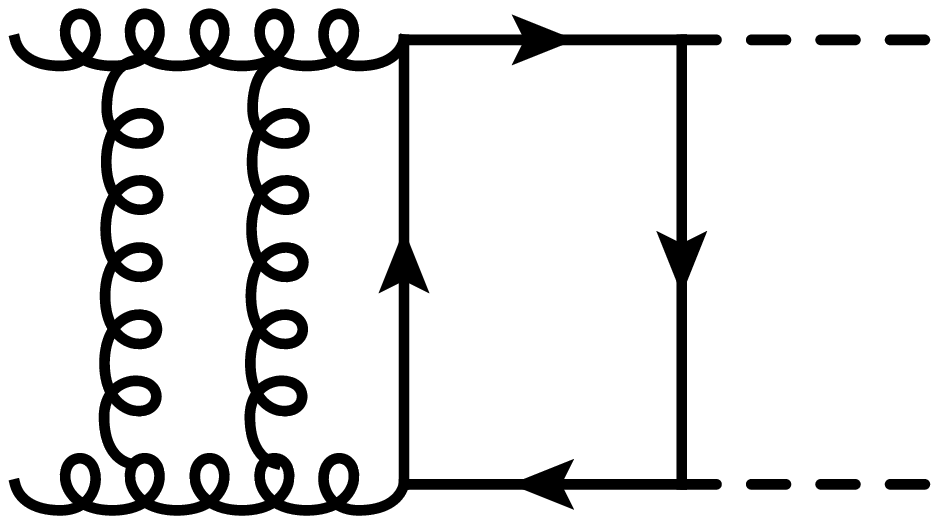}}
    \\
    $(Tn_h)^2$ & $C_F^2Tn_h$ & $C_AC_FTn_h$ & $C_A^2Tn_h$
    \\
    \raisebox{1.2em}{\includegraphics[width=0.18\textwidth]{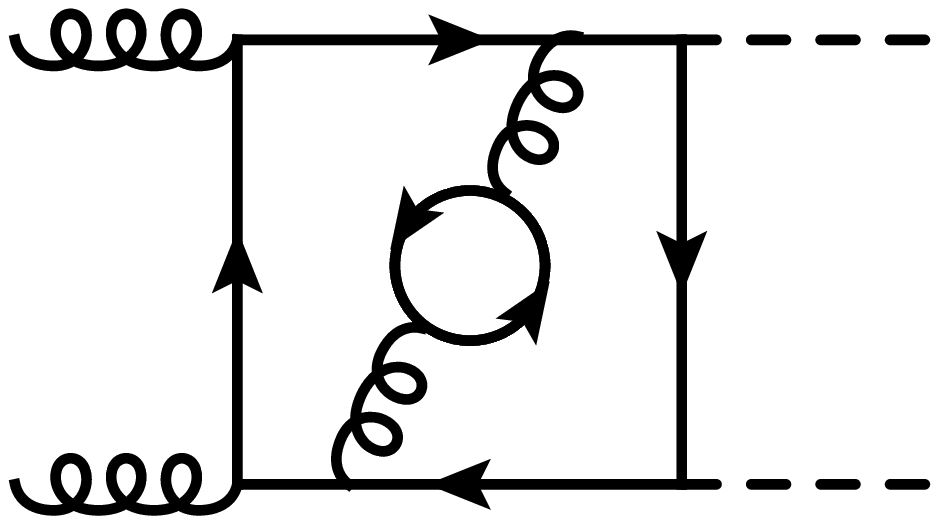}}  &
    \raisebox{1.2em}{\includegraphics[width=0.18\textwidth]{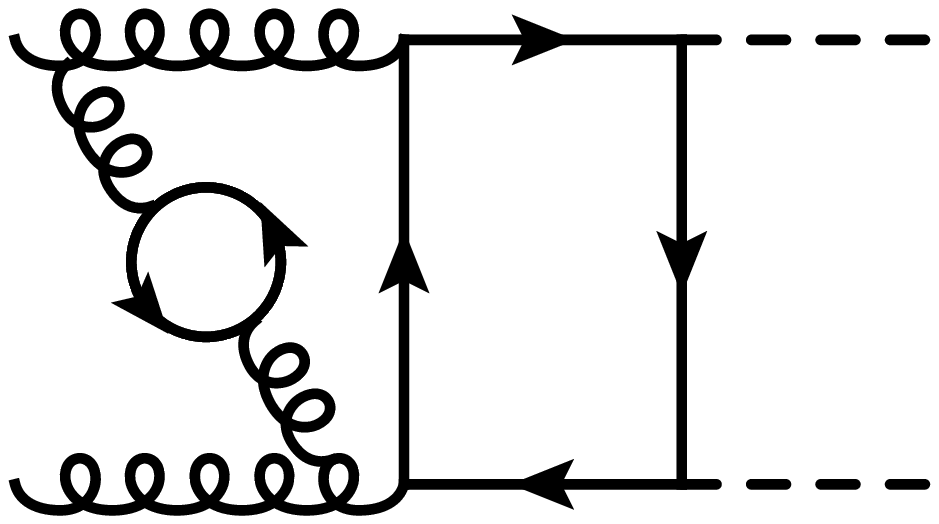}}  &
    \includegraphics[width=0.15\textwidth]{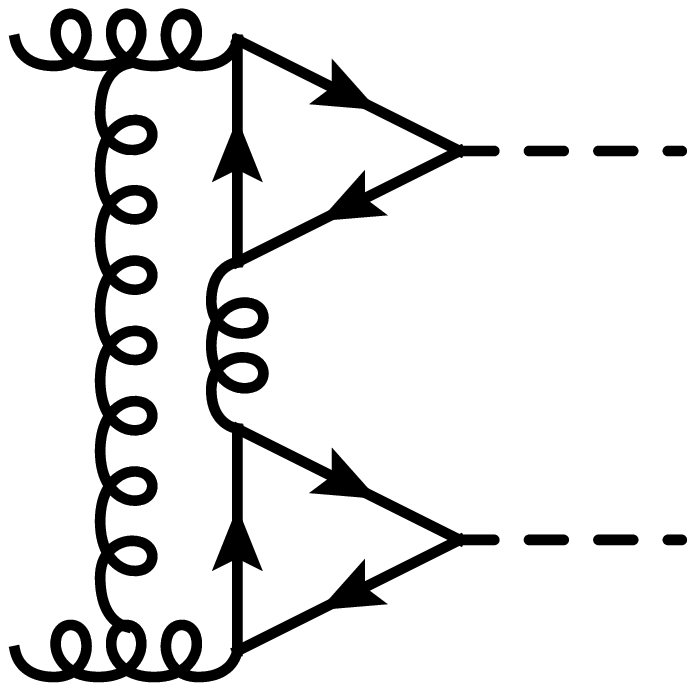} &
    \includegraphics[width=0.15\textwidth]{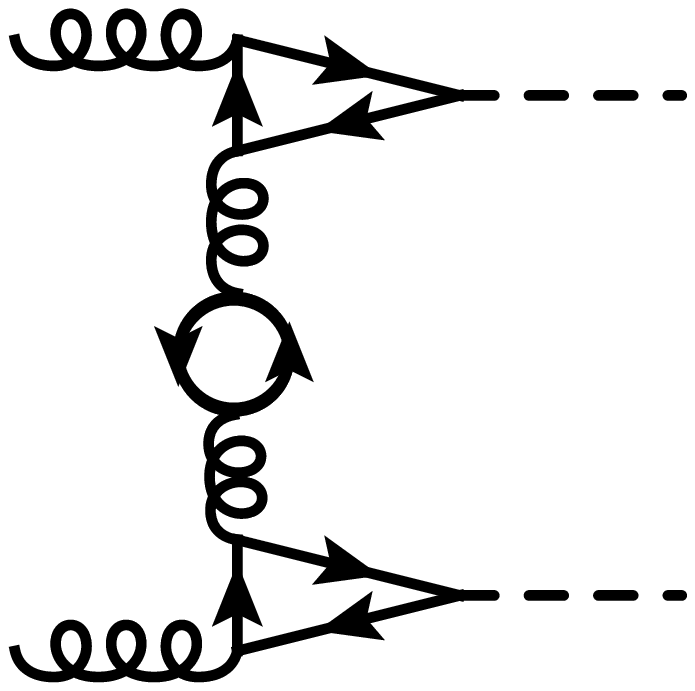}
    \\
    $C_F(Tn_h)^2$ and &
    $C_A(Tn_h)^2$ and &
    $C_A(Tn_h)^2$ &
    $(Tn_h)^3$
    \\
    $C_FT^2n_hn_l$ &
    $C_AT^2n_hn_l$
  \end{tabular}
  \caption{\label{fig::diags}Sample Feynman diagrams contributing to
    $gg\to HH$. For simplicity we show diagrams with a triple-Higgs boson
    coupling only at one-loop order. A sample colour factor is shown
      below each diagram. However, note that in general a diagram contributes to more
      than one colour structure.  Solid, dashed and curly lines denote
    quarks, Higgs bosons and gluons respectively.}
\end{figure}

We generate the Feynman amplitudes with the help of {\tt
  qgraf}~\cite{Nogueira:1991ex} and obtain 11, 197 and 5703 diagrams at one,
two and three loops. Note that both one-particle irreducible (1PI) and
one-particle reducible (1PR) contributions have to be considered. Sample
diagrams are shown in Fig.~\ref{fig::diags} together with the corresponding
colour factors expressed in terms of the Casimir invariants of SU$(N_c)$:
$C_A=N_c$ and $C_F=(N_c^2-1)/(2N_c)$.  Furthermore we have $T=1/2$ and use
the labels $n_l$ and $n_h$ for closed massless and massive fermion loops respectively.
For numerical evaluation we set $n_l=5$ and $n_h=1$.  In the following subsections
we provide several technical details of the calculation of the form factors.

\subsection{\label{sec::ae}Asymptotic expansion}

The programs {\tt q2e} and {\tt
  exp}~\cite{Harlander:1997zb,Seidensticker:1999bb,q2eexp} have been designed
to work hand-in-hand when applying a (possibly nested) asymptotic expansion
involving a large external momentum or a large internal mass to an amplitude
generated by {\tt qgraf}~\cite{Nogueira:1991ex}. The output of {\tt exp} is
{\tt FORM}~\cite{Ruijl:2017dtg} code\footnote{In the computations
  described in this paper we mainly use the parallel version, {\tt TFORM}.} for
each sub-diagram which has to be considered according to the rules of
asymptotic expansion (see, e.g., Ref.~\cite{Smirnov:2012gma}).

In this case we apply the rules of asymptotic expansion for the limit
\begin{eqnarray}
  m_t \gg q_1, q_2, q_3\,,
  \label{eq::limit}
\end{eqnarray}
where $q_1^2=q_2^2=0$ are the incoming gluon momenta and
$q_3^2=m_H^2$. Equation~(\ref{eq::limit}) implies that the Feynman
amplitudes are expanded in powers of
\begin{eqnarray}
  \{q_3\cdot q_3,\:\: q_1\cdot q_2,\:\: q_1\cdot q_3,\:\: q_2\cdot q_3\}/m_t^2,
\end{eqnarray}
possibly multiplied by logarithms of these ratios.

The main purpose of Eq.~(\ref{eq::limit}) is the reduction of the number of scales
in the loop integrals. Furthermore, the three-loop integrals are
factorized into products of lower-loop integrals.  In the box diagrams we initially
have the scales $s$, $t$, $m_H^2$ and $m_t^2$ and in the triangle diagrams
$s$ and $m_t^2$. After asymptotic expansion we find the following
products of integrals
\begin{center}
  \begin{tabular}{ll|l}
    \hline
    \multicolumn{2}{c|}{Type of integrals for} & scales \\ 
    hard subgraph & \hphantom{$\times$ }co-subgraph & \\
    \hline
    3-loop vacuum   & --- & $m_t^2$ \\ 
    2-loop vacuum   &$\times$ 1-loop massless triangle & $m_t^2 \times s$\\
    two 1-loop vacuum &$\times$ 1-loop massless box      & $m_t^2 \times s,t,m_H^2$\\
    1-loop vacuum   &$\times$ 2-loop massless triangle & $m_t^2 \times s$\\
    \hline
  \end{tabular}
\end{center}
Note that integrals with more than one scale only have to be considered at
one-loop order; the corresponding integral families are well-studied in the
literature~\cite{Birthwright:2004kk,Ellis:2007qk,Chavez:2012kn}.
In the above table ``massless'' refers to the propagator masses only.
Dependence on the Higgs boson mass is retained.
In the one-loop massless box case, degenerate cases also occur for which
one of the scales is absent.

In cases in which one has to deal with products of integrals we organize the
output of {\tt exp} in such a way that we perform the vacuum integrals first,
since it is simpler to compute vacuum tensor integrals than tensor integrals
for families with external momenta. In fact, at one and two loops vacuum
tensor integrals with arbitrary rank can be treated.\footnote{A closed formula
  for the one-loop case can, e.g., be found in~\cite{Smirnov:2012gma} and an
  algorithm for the two-loop case is presented in~\cite{Chetyrkin:1993rv}.}
For three-loop vacuum integrals we implement projectors which are discussed in
detail in Subsection~\ref{sec::proj}.  For the remaining massless integrations
no tensor integrals have to be solved.

The vacuum integrals are performed with the {\tt FORM} package {\tt
  MATAD}~\cite{Steinhauser:2000ry} and for the massless integral families we use
{\tt FIRE}~\cite{Smirnov:2014hma} to obtain integral tables which
express all integrals appearing in the amplitudes in terms of master integrals (see
Fig.~7 of Ref.~\cite{Grigo:2015dia} for graphical
representations). Analytic expressions for the latter can be found in
Refs.~\cite{Birthwright:2004kk,Ellis:2007qk,Chavez:2012kn,Grigo:2015dia}.

\begin{figure}[t]
  \centering
  \begin{tabular}{c|c}
    \includegraphics[width=0.35\textwidth]{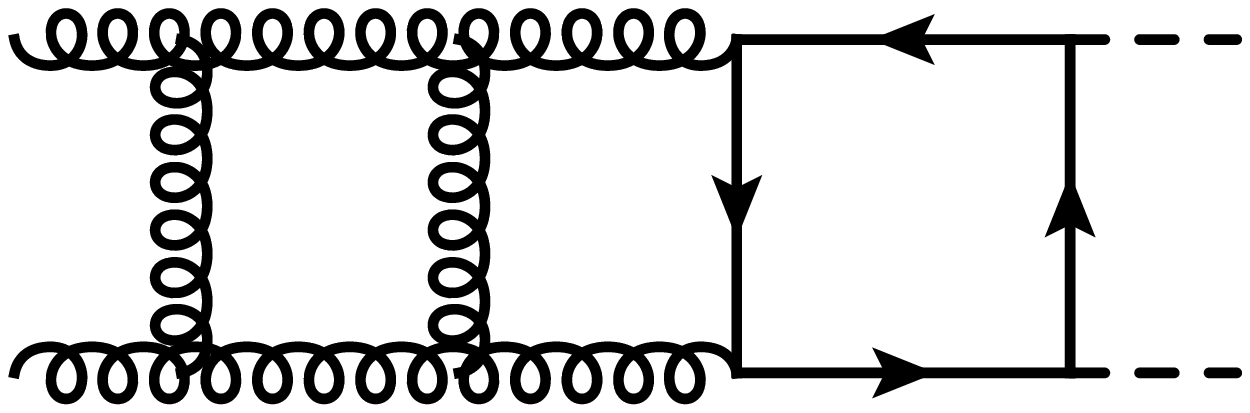}
    &
      \includegraphics[width=0.20\textwidth]{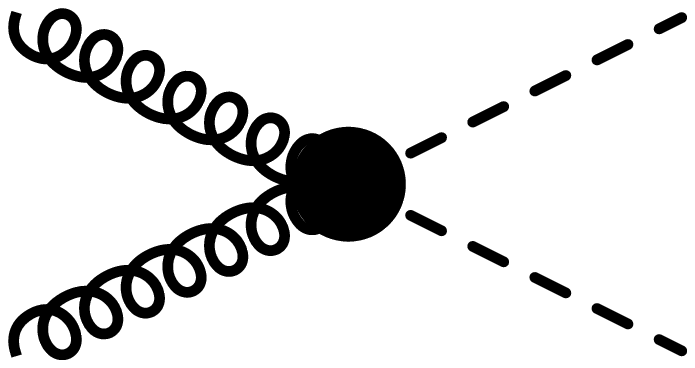}
      \includegraphics[width=0.25\textwidth]{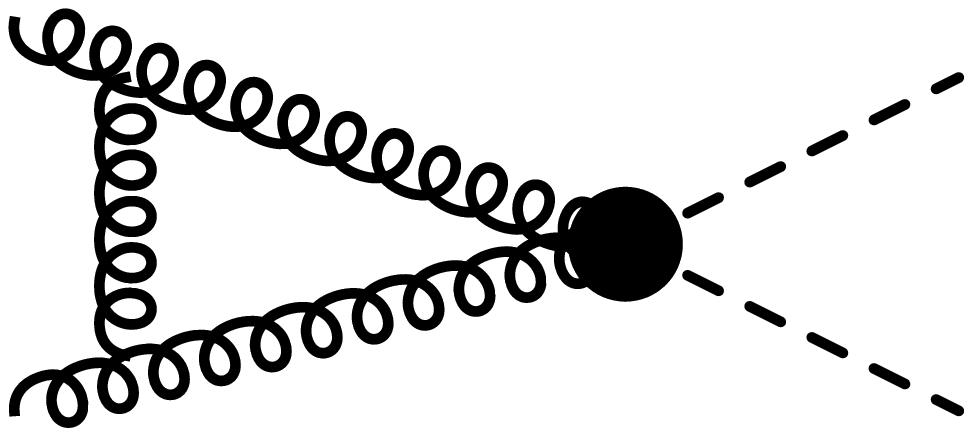}
    \\ &
         \includegraphics[width=0.25\textwidth]{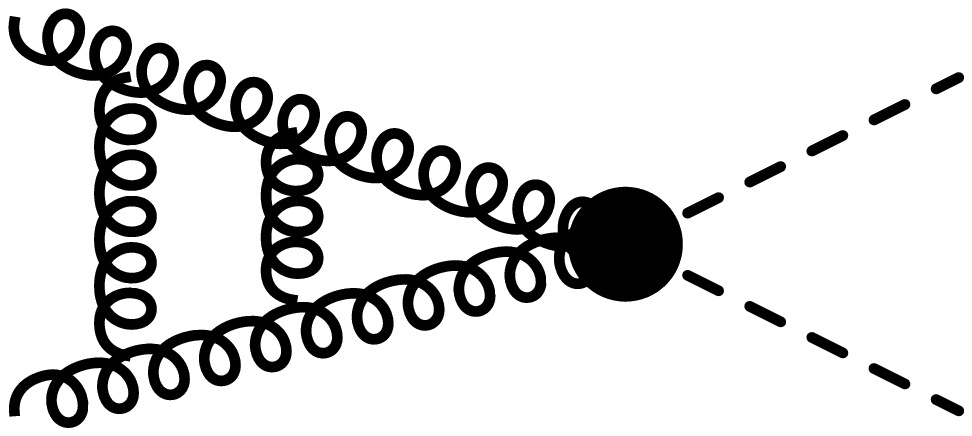}
    \\ \hline \hline \\
    \includegraphics[width=0.25\textwidth]{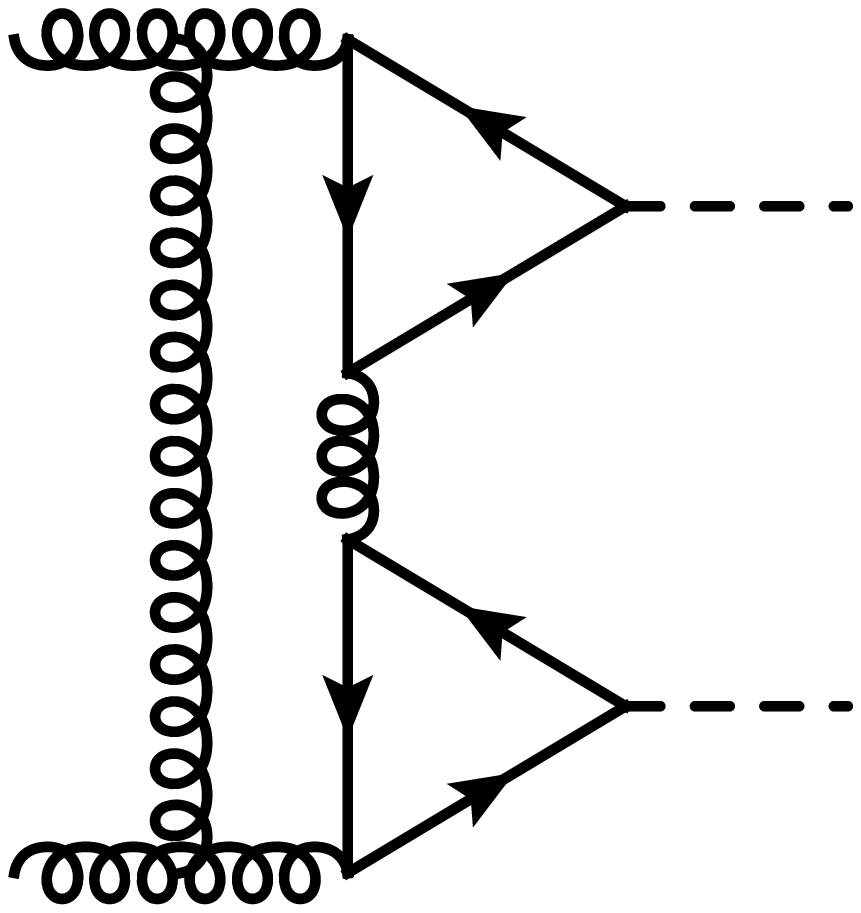}
    &\raisebox{3em}{
      \includegraphics[width=0.20\textwidth]{gghhFF3la_c3.eps}
      \includegraphics[width=0.15\textwidth]{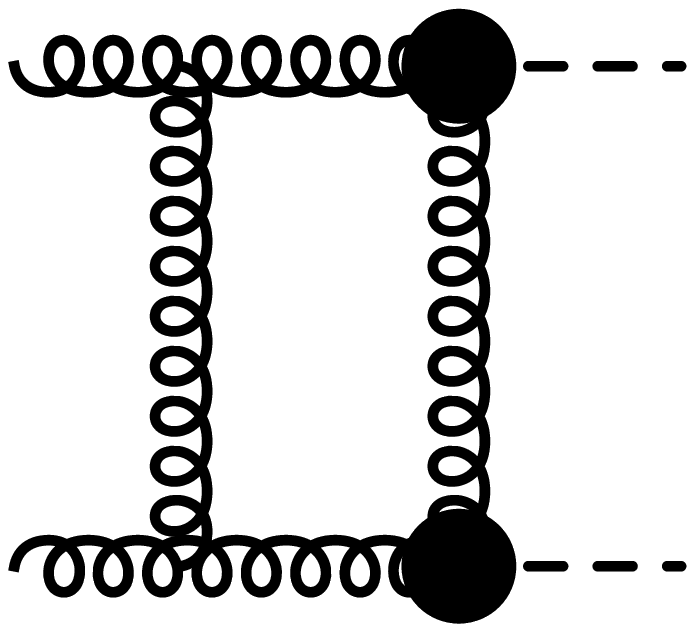}
      }
    \end{tabular}
    \caption{\label{fig::diags_ae}Sample three-loop diagrams (left) and the
      corresponding co-subgraphs (right) which result from the asymptotic expansion
      according to Eq.~(\ref{eq::limit}). The blobs represent effective
      vertices from the hard subgraphs. They correspond to vacuum integrals.}
\end{figure}

Let us illustrate the procedure described above using two typical Feynman
diagrams shown in Fig.~\ref{fig::diags_ae}. We show the three-loop
diagrams which have to be expanded in all external momenta, and the corresponding lower-loop
co-subgraphs which appear after applying the scale hierarchy of
Eq.~(\ref{eq::limit}). The blobs represent effective vertices from the expanded
hard subgraphs which we do not show explicitly.

Note that due to the rules of asymptotic expansion the hard subgraphs
have to expanded in all small quantities, which in this case are the
external momenta $q_i$ but also the loop momenta of the co-subgraphs. This
results in a multi-dimensional Taylor expansion which we want to compute
up to $5^{\rm th}$ order (i.e. up to order $1/(\mt^2)^4$) for the box form
factors and up to $8^{\rm th}$ order ($1/(m_t^2)^7$) for the triangle form factor.
At this point the intermediate expressions can become huge and special
measures and optimizations have to be applied in order to obtain the results
with the computing resources available.
These methods are described in the following subsection.

\subsection{\label{sec::proj}Projection}

A major bottleneck in the computation of~\cite{Grigo:2015dia} is the
calculation of three-loop tensor vacuum integrals. After expansion in
$1/\mts$ the intermediate expressions become rather large, which cause these
routines to perform very poorly. In order to avoid this issue in this work we
project the sub-diagrams which contain a three-loop vacuum integral onto a
suitable ansatz, and compute individual terms of this ansatz separately. The
intermediate expressions for each term become much smaller, and we no longer
have to compute tensor integrals.
The diagrams contributing to the triangle form factor have a simpler structure,
and thus use a simplified version of the method discussed below. For this
reason we are able to compute an additional three terms in the expansion,
compared to the depth of the box-type form factors.
We elaborate on this at the end of the subsection.

Each diagram can be written in the following way, (see
also~\cite{Fleischer:1994ef}, here we extend the ansatz to account for the
additional external momentum),
\begin{equation}
\label{eqn:ExpansionAnsatz}
  A = \sum_{L=0}^{L_{max}} \:\: \sum_{i+j+k+l+m+n=L} C_{i,j,k,l,m,n}\: (q_1^2)^i\: (q_2^2)^j\:
  (q_3^2)^k\: (q_1 \cdot q_2)^l\: (q_1 \cdot q_3)^m\: (q_2 \cdot q_3)^n.
\end{equation}
where $L_{max}$ depends on the depth of the $1/\mts$ expansion being considered.
Since we consider the process $g(q_1)g(q_2)\rightarrow H(q_3)H(q_4)$ we have
that $q_1^2 = q_2^2 = 0$; we can therefore set $i=j=0$ in the ansatz
here. Associated with the six possible scalar products between the momenta are
six derivative operators
\begin{equation}
  \square_{a,b} = \frac{\partial}{\partial q_{a\:\mu}} \frac{\partial}{\partial q_b^{\:\mu}},
\end{equation}
with which one can construct projection operators $P_{i,j,k,l,m,n}$ to project
particular coefficients $C_{i,j,k,l,m,n}$ of the ansatz $A$ from the amplitude, i.e.
\begin{equation}
	P_{i,j,k,l,m,n}\: A = C_{i,j,k,l,m,n}.
\end{equation}
It is understood that after taking the derivatives contained in the projector
terms, all remaining external momenta of the diagram are set to zero. The
first few projection operators are as follows, where we define the notation
$\square_{i,j,k,l,m,n} = \square_{1,1}^{\:\:i} \square_{2,2}^{\:\:j}
\square_{3,3}^{\:\:k} \square_{1,2}^{\:\:l} \square_{1,3}^{\:\:m}
\square_{2,3}^{\:\:n}$ and as above, $d = 4-2\epsilon$,
\begin{align}
\label{eqn:AnsatzProjectors}
L=1:\hspace{1cm}
\nonumber\\
	P_{0,0,0,0,0,1} &=
		\square_{0,0,0,0,0,1}\: \frac{1}{d},
\qquad
	P_{0,0,0,0,1,0} =
		\square_{0,0,0,0,1,0}\: \frac{1}{d},
\nonumber\\
	P_{0,0,0,1,0,0} &=
		\square_{0,0,0,1,0,0}\: \frac{1}{d},
\qquad
	P_{0,0,1,0,0,0} =
		\square_{0,0,1,0,0,0}\: \frac{1}{2d},
\nonumber\\[2mm]
L=2:\hspace{1cm}
\nonumber\\
	P_{0,0,0,0,0,2} &=
		\square_{0,0,0,0,0,2}\: \frac{1}{2 d^2 + 2 d - 4}
		+ \square_{0,1,1,0,0,0}\: \frac{-1}{2 d^3 + 2 d^2 - 4 d},
\nonumber\\
	P_{0,0,0,0,1,1} &=
		\square_{0,0,0,0,1,1}\: \frac{1}{d^2 + d - 2}
		+ \square_{0,0,1,1,0,0}\: \frac{-1}{d^3 + d^2 - 2 d},
\nonumber\\
&\vdots
\nonumber\\
	P_{0,0,1,1,0,0} &=
		\square_{0,0,0,0,1,1}\: \frac{-1}{d^3 + d^2 - 2 d}
		+ \square_{0,0,1,1,0,0}\: \frac{d + 1}{2 d^3 + 2 d^2 - 4 d},
\nonumber\\
	P_{0,0,2,0,0,0} &=
		\square_{0,0,2,0,0,0}\: \frac{1}{8 d^2 + 16 d}.
\end{align}
For the $1/\mt^8$ terms we need such projection operators at $L=6$.  This is
because the vertex diagrams have mass dimension two which are built from
combinations of external momenta as required by gauge invariance. Note that
contributions involving $\square_{1,1}$ and $\square_{2,2}$ are needed in
the construction of the projection operators even though $i=j=0$ is chosen in
Eq.~(\ref{eqn:ExpansionAnsatz}).  

To compute these projections efficiently, we form linear combinations of
the projection operators which involve just a single derivative operator
$\square_{i,j,k,l,m,n}$. For example at $L=2$, $\square_{0,0,1,1,0,0}$ is
present in $P_{0,0,0,0,1,1}$ and $P_{0,0,1,1,0,0}$. Thus, one obtains
contributions to the $\left(q_1\cdot q_3\right) \left(q_2\cdot q_3\right)$
and $\left(q_3\cdot q_3\right) \left(q_1\cdot q_2\right)$ terms of the
expansion ansatz by applying the linear combination
\begin{align}
      \label{eqn:diffproj2}
      & \bigg[
        \bigg(
        -\frac{1}{72}
        -\frac{1}{48}\ep
        -\frac{55}{2592}\ep^2
        -\frac{95}{5184}\ep^3
        +\mathcal{O}(\ep^4)
        \bigg) \left(q_1\cdot q_3\right) \left(q_2\cdot q_3\right) +
        \nonumber\\
      & \hphantom{\bigg[}\bigg(
        \frac{5}{144}
        +\frac{11}{288}\ep
        +\frac{167}{5184}\ep^2
        +\frac{85}{3456}\ep^3
        +\mathcal{O}(\ep^4)
        \bigg) \left(q_3\cdot q_3\right) \left(q_1\cdot q_2\right)
        \bigg]\: \square_{0,0,1,1,0,0}
    \end{align}
to the amplitude. Here the rational polynomials in $d$ have been expanded.

We compute all necessary derivative operators applied to the diagrams after the expansion in
$1/\mts$, before we perform the three-loop vacuum integral procedures. Each derivative operator
(that is, each $\square_{i,j,k,l,m,n}$ required) is applied as a separate task and all
results are combined at the end. This ensures that intermediate expressions remain a
manageable size, and that no derivative operator is computed more than once.

For reasonable performance it is crucial to not repeat the $1/\mts$ expansion
of the diagrams for each of the above tasks, since it is a very computationally
expensive procedure. The expansion is performed just once; the intermediate result
is then split into parts containing particular numbers of each external momentum
and stored. The projection tasks can load exactly the part which will yield a
non-zero result after taking the derivatives with respect to the external momenta.

The structure of the computation is summarized below. For some aspects we
provide a more detailed description in Section~\ref{sec::optimizations}.

\vspace{1em}

\begin{enumerate}
\item \textbf{$1/\mts$ expansion:}
  \begin{enumerate}
  \item Sum all diagrams with the same colour factor to make
    ``super-diagrams''. Many terms are common to multiple diagrams, so
    summing them reduces the total size of the intermediate
    expressions. At three loops there are 5703 Feynman diagrams which form nine
    super-diagrams with the colour factors (considering only three-loop
    vacuum sub-diagrams),
    \begin{align}
      &d_{abc}d^{abc} \nh^2,\quad
        d_{abc}d^{abc} \nh\nl,\quad
        \ca T^2\nh^2,\quad
        \cf T^2\nh^2,\quad
        \nonumber\\
      &\ca^2 T\nh,\quad
        \ca\cf T\nh,\quad
        \cf^2 T\nh,\quad
        \ca T^2\nh\nl,\quad
        \cf T^2\nh\nl.
    \end{align}
    The super-diagrams with colour factors proportional to $d_{abc}d^{abc}$,
    which arise from Feynman diagrams with two closed fermion loops with three
    gluon couplings each, are found to vanish after expansion in $1/\mts$ in
    Step~1.~(d) (see below), which is why this
    colour structure is not listed in Fig.~\ref{fig::diags}.  Note that of
    the eight three-loop colour structures listed in Fig.~\ref{fig::diags}
    only $(Tn_h)^3$ has no 1PI three-loop vacuum contribution.
    
  \item For each super-diagram, multiply by one of the five Lorentz structures
    of the amplitude projectors (c.f. Eq.~(\ref{eqn:FFprojectors})),
    \begin{equation}
      q_{1,\nu} q_{2,\mu},\quad
      q_{1,\nu} q_{3,\mu},\quad
      q_{2,\mu} q_{3,\nu},\quad
      q_{3,\mu} q_{3,\nu},\quad
      g_{\mu\nu} q_{12}.
    \end{equation}
    This produces $5\times 9 = 45$ projected super-diagrams, to be expanded in
    $1/\mts$. Apply Feynman rules and perform Dirac algebra. The coefficients
    of these Lorentz structures (in the round brackets of
    Eq.~(\ref{eqn:FFprojectors})) will be included when everything is combined
    in Step~2.~(b).
  \item Use graph symmetries to reduce the number of terms and size of
    expressions.
  \item Perform the $1/\mts$ expansions. These are heavy computations, for
    which we use computing nodes with relatively large amounts of memory and
    processing cores (at least 96GB memory and 12 cores). It is crucial to not
    duplicate any work here; we make extensive use of the \texttt{FORM}
    statements \texttt{Collect} (to reduce the number of terms to be
    processed) and \texttt{ArgToExtraSymbol} (to temporarily reduce the size
    of the expressions).
    After expansion, graph symmetries are again used to reduce the number of
    terms and size of the expressions.

	The five most difficult projected super-diagrams are those with colour factor $C_A^2 T n_h$.
	To expand to $1/\mt^8$ these each require a wall time of around 10 days.
	The total size of the (\texttt{gzip} compressed) stored expressions
    for the expansions of the 45 projected super-diagrams is 324GB.
  \end{enumerate}
  
\item \textbf{Projection:}
  \begin{enumerate}
  \item For each of the necessary operators (see Eq.~(\ref{eqn:diffproj2})), load the
    relevant part of the expanded
    super-diagram (for the example of Eq.~(\ref{eqn:diffproj2}), the part
    containing terms with one $q_1$, one $q_2$ and two $q_3$). All other parts
    would yield zero after differentiation, so do not need to be loaded.

    The differentiation must be performed inside \texttt{FORM} \texttt{CFunction}
    arguments to avoid an enormous blow-up of intermediate expression
    sizes. These tasks are much easier, computationally, than those of the
    expansion steps. They are run requiring only 8GB of memory and 4
    processing cores.
    To obtain the $1/\mt^{\{0,2,4,6,8\}}$ terms of the expansion there are
    \{15, 38, 88, 174, 324\} derivatives to compute for each of the 45 projected
    super-diagrams, yielding \{675, 1,710, 3,960, 7,830, 14,580\} tasks to be run respectively.
    These tasks required a total time of approximately 1,600 days to complete; running
    tasks concurrently on our cluster this corresponds to a wall time of about 1 month.

  \item The results of these operators applied to the diagrams allow one to construct
    the result in the form of the ansatz Eq.~(\ref{eqn:ExpansionAnsatz}). Combining all terms,
    along with the coefficients of the Lorentz structures of Eq.~(\ref{eqn:FFprojectors}),
    yields the final result for the form factors ${\cal M}_1$ and ${\cal M}_2$.
  \end{enumerate}
\end{enumerate}

As mentioned above, some simplifications are possible when computing the
triangle form factor. It comes only with the Lorentz structures $g_{\mu\nu}$
and $q_{1,\nu}q_{2,\mu}$ (see Eq.~(\ref{eq::A1A2})), thus in step 1.~(b) fewer projected
super-diagrams are produced since we can ignore the additional three structures
required by the box-type form factors. The ansatz of Eq.~(\ref{eqn:ExpansionAnsatz}) can also be
simplified; only the index $l$ needs to be non-zero. Thus, fewer derivative operators
need to be computed in step 2.~(a): for the $1/m_t^{\{0,2,4,6,8,10,12,14\}}$
terms of the expansion we must apply just $\{2,2,3,3,4,4,5,5\}$
derivative operators.

\subsection{\label{sec::optimizations}Calculation optimizations}
In this Section we outline a few methods by which we were able to optimize the computation, in
addition to the projection procedure described above. We note that without such optimizations,
computing the expansion to a depth $1/\mt^8$ (and likely $1/\mt^6$) for
the box-type and $1/\mt^{14}$ for the triangle form factors would not have been possible
with the computing resources available to us.

\subsubsection{Graph Symmetries}
In Step 1.~(c) and 1.~(d) we use graph symmetries to reduce the size of the
intermediate expressions. We map the vacuum integrals to a minimal set by
using rotation and reflection symmetries, implemented by re-labelling the line
momenta of equivalent graphs such that they coincide. Some examples of this
procedure are shown in Table~\ref{tab::graphsym}.

After expansion in $1/\mts$ many integrals appear with higher-power
(``dotted'') propagators. One can move the dots around the graph, using the
same symmetry relations as described above, to obtain a smaller set of
integrals.

\begin{table}[b]
\centering
\begin{tabular}{|c|c|c|c|}
\hline 
Top-level Topology & Graph 1 & Graph 2 & Relabelling \\ 
\hline 
	\includegraphics[scale=0.8]{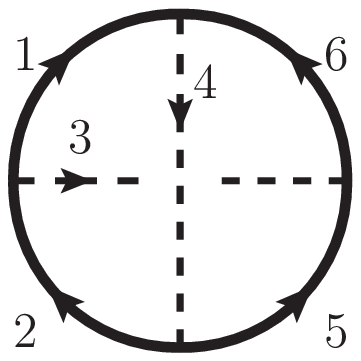} &
	\includegraphics[scale=0.8]{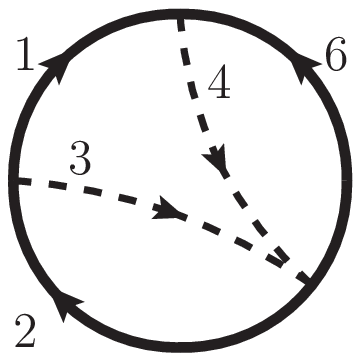} &
	\includegraphics[scale=0.8]{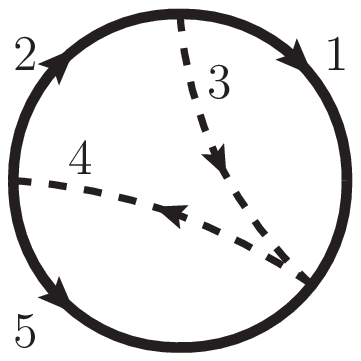} &
	\begin{minipage}[b]{2cm}
	\begin{align*}
		p1 \rightarrow \hphantom{-}p2\\
		p2 \rightarrow -p5\\
		p3 \rightarrow -p4\\
		p4 \rightarrow \hphantom{-}p3\\
		p6 \rightarrow -p1
	\end{align*}
	\end{minipage}\\
\cline{2-4}
	&
	\includegraphics[scale=0.8]{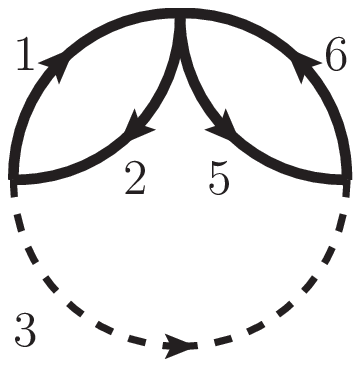} &
	\includegraphics[scale=0.8]{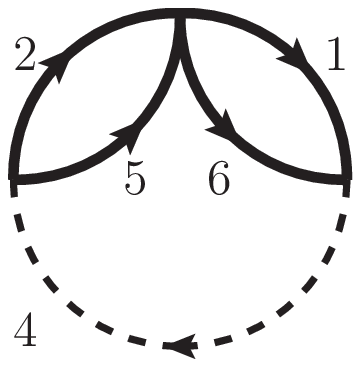} &
	\begin{minipage}[b]{1cm}
	\begin{align*}
		p1 \rightarrow \hphantom{-}p2\\
		p2 \rightarrow -p5\\
		p3 \rightarrow -p4\\
		p5 \rightarrow \hphantom{-}p6\\
		p6 \rightarrow -p1
	\end{align*}
	\end{minipage}\\
\hline
\end{tabular}
\caption{\label{tab::graphsym} Graphs 1 and 2 are derived from the Top-level Topology, with different lines missing. This yields different, but equivalent, graphs. Line momenta are relabelled to make this equivalence manifest; we show the replacements required to map Graph 1 onto Graph 2. The arrows denote the direction of momentum flow.}
\end{table} 

\subsubsection{ArgToExtraSymbol}
In step 1.~(d), the $1/\mts$ expansions are performed. At this point, the
\texttt{FORM} representation of the terms in the expressions looks something
like
\begin{verbatim}
+ Den(l1,mt) * Den(l1+q1,mt) * ... * Den(l2-q3,mt) * ( many terms )
\end{verbatim}
where the \texttt{Den} functions represent the propagators to be expanded;
they are of the form $1/(m_t^2 - (l_1+q_1)^2)$, for example.  The
``\texttt{many terms}'' inside the brackets are coefficients which do not take
part in the expansion. Since there can be many thousand such coefficients,
it is crucial to keep them bracketed away during the multi-module expansion
routine, to keep the number of terms small and avoid expanding the same
product of \texttt{Den} functions many times. One typically achieves this with
a construction such as
\begin{verbatim}
Bracket Den;
.sort
CFunction f;
Collect f;
\end{verbatim}
which moves the bracketed terms inside the argument of \texttt{f}. While this does
indeed keep the number of terms small, it does not (greatly) reduce the
size of the expression. If the expression is large enough to require
disk-based sorting at the end of each module of the expansion routine, one
still has a severe performance bottleneck. We resolve this by additionally
making use of the statement \texttt{ArgToExtraSymbol f;} after \texttt{Collect f;},
which replaces the (large) arguments of the \texttt{f}s with unique symbols, whose
definitions are stored by \texttt{FORM}. More memory is required to store these
definitions, but the resulting reduction in size of the expression involved in
disk-based sorting provides a large speed-up of the expansion routine. After
expansion is complete the original coefficients may be recovered with the
\texttt{FromPolynomial} statement. 

Let us remark that the use of \texttt{ArgToExtraSymbol} is also
essential to make possible and speed up the calculation of the
subdiagrams where two-loop vacuum tensor integrals are needed.

\subsubsection{Compression}
In step 1.~(d) it was stated that the intermediate results of the $1/\mts$
expansions are compressed with \texttt{gzip} and stored, for use in step
2.~(a). Since these compressed results occupy 324GB, they cannot easily be
stored uncompressed on the storage available to us. In step 2.~(a) the tasks
can easily retrieve the relevant compressed intermediate result
from network storage and decompress it onto the local storage of the node on
which they are running, by making use of \texttt{FORM}'s \texttt{\#system}
preprocessor command:
\begin{verbatim}
#system gunzip < /network/intermediate.sav.gz > /local/intermediate.sav
Load /local/intermediate.sav;
...
\end{verbatim}
As well as reducing the capacity required for the storage of the intermediate
results, this also provides a large performance improvement by reducing the
I/O load of the network and storage hardware when hundreds of tasks are running
concurrently.


\section{\label{sec::ren}Renormalization and infrared subtraction}


\subsection{Ultraviolet divergences}

The renormalization of the ultraviolet (UV) divergences is
straightforward:
\begin{itemize}
\item The top quark mass ($m_t$) renormalization is needed up to two
  loops. We first renormalize $m_t$ in the $\overline{\rm
    MS}$ scheme, and then transform $m_t$ to the on-shell
  scheme. Note that higher order $\epsilon$ terms are needed in the
  corresponding one-loop expression since the NLO (two-loop) amplitude
  develops $1/\epsilon^2$ poles, even after all UV counter-terms are taken
  into account.  Since the LO (one-loop) amplitude is finite the
  two-loop term in the $\overline{\rm MS}$-on-shell conversion formula
  is only needed up to ${\cal O}(\epsilon^0)$.

\item The gluon wave function renormalization is also needed up to two
  loops. Note that, since we work in dimensional regularization, where
  scaleless integrals are set to zero, only diagrams with virtual top
  quarks contribute. These two-point functions have to be computed for
  $q^2=0$ which corresponds to on-shell gluons.  Note that the gluon
  wave function renormalization agrees with the decoupling constant of
  the gluon field needed to relate five- and six-flavour
  QCD~\cite{Chetyrkin:1997un}.

\item The strong coupling constant renormalization up to two loops is
  performed in full six-flavour theory.

\item Finally the decoupling relation for $\alpha_s$ is needed up
  to two loops in order to express $\alpha_s^{(6)}$ in terms of
  $\alpha_s^{(5)}$.  Similar to the $\overline{\rm MS}$--on-shell mass
  relation also here the one-loop expression is needed up to order
  $\epsilon^2$.
\end{itemize}
  
The final result is expressed in terms of the top quark pole mass,
and the five-flavour strong coupling, $\alpha_s^{(5)}$.  It
still contains poles up to order $1/\epsilon^4$ which are of infrared
nature.  They will be treated in the next subsection.


\subsection{Subtraction of infrared divergences}

For the subtraction of the infrared (IR) poles we follow
Ref.~\cite{Catani:1998bh}, see also
Refs.~\cite{deFlorian:2012za,Grigo:2014jma}. For convenience we
provide explicit expressions for the subtraction terms.  We apply the
IR subtraction at the amplitude level since we want the obtain
finite expressions for the form factors.

After UV renormalization we have the following colour factors
at one-, two- and three-loop order:
\begin{eqnarray}
  &&T n_h
     \,,\nonumber\\
  &&T n_h \{C_F, C_A, Tn_h, Tn_l\}
     \,,\nonumber\\
  &&T n_h \{C_F^2, C_AC_F, C_A^2, 
     C_FTn_l, C_ATn_l, 
     C_FTn_h, C_ATn_h, 
     T^2n_l^2, T^2n_hn_l, T^2n_h^2\}.
     \label{eq::colfac}
\end{eqnarray}
In the following discussion we omit the overall factor $T n_h$
which is contained in the quantity $X_0$, see Eq.~(\ref{eq::X0}).
Note that the structures $T^2n_hn_l$, $T^3n_hn_l^2$ and $T^3n_h^2n_l$
are not present in the two- and three-loop diagrams (cf. Fig.~\ref{fig::diags})
but only arise from UV counter-terms and IR subtraction (see below).

After UV renormalization,
at two-loop order the colour factors $\{C_A, Tn_l\}$ come with
$1/\epsilon$ poles and $C_A$ also has a $1/\epsilon^2$ pole.
At three-loop order, highest-order pole appearing with each colour factor is summarized in the
following table,
\begin{center}
\begin{tabular}{|c|l|}
\hline 
Leading Pole & Colour Factors \\ 
\hline 
$1/\epsilon^4$ & $C_A^2$ \\ 
\hline 
$1/\epsilon^3$ & $C_A T n_l$ \\ 
\hline 
$1/\epsilon^2$ & $C_A C_F$, $C_A T n_h$, $T^2 n_l^2$ \\ 
\hline 
$1/\epsilon$ & $C_F T n_l$, $T^2 n_h n_l$\\ 
\hline 
$1$ & $C_F^2$, $C_F T n_h$, $T^2 n_h^2$ \\ 
\hline 
\end{tabular}
\end{center}
We have checked that all these poles cancel after applying the following IR
subtraction procedure: finite form factors, $F^{\rm fin}_X$, at NLO and NNLO
are obtained via
\begin{eqnarray}
  F^{(1),\rm fin}_X &=& F^{(1)}_X - \frac{1}{2} I^{(1)}_g F^{(0)}_X\,,\nonumber\\
  F^{(2),\rm fin}_X &=& F^{(2)}_X - \frac{1}{2} I^{(1)}_g F^{(1)}_X
  - \frac{1}{4} I^{(2)}_g F^{(0)}_X\,,
  \label{eq::FF_IR}
\end{eqnarray}
where, as in Eq.~(\ref{eq::F}), $X\in\{\mbox{tri},\mbox{box1},\mbox{box2}\}$.
The quantities on the r.h.s. of Eq.~(\ref{eq::FF_IR})
are UV-renormalized.
$I^{(1)}_g$ and $I^{(2)}_g$ on the r.h.s. of Eq.~(\ref{eq::FF_IR})
are given by~\cite{Catani:1998bh,deFlorian:2012za}
\begin{eqnarray}
	I^{(1)}_g &=&
		{} - \left(\frac{\mu^2}{-s-i\delta}\right)^\epsilon
		\frac{e^{\epsilon\gamma_E}}{\Gamma(1-\epsilon)}
		\frac{1}{\epsilon^2}
		\Big[
			C_A + 2\epsilon\beta_0
		\Big]\,,\\
	I^{(2)}_g &=&
		{} - \left(\frac{\mu^2}{-s-i\delta}\right)^{2\epsilon}
		\left(\frac{e^{\epsilon\gamma_E}}{\Gamma(1-\epsilon)}\right)^2
		\frac{1}{\epsilon^4}
		\Big[
			\frac{1}{2}(C_A+2\epsilon\beta_0)^2
		\Big]\nonumber\\
		&& {} + \left(\frac{\mu^2}{-s-i\delta}\right)^{\epsilon}
		\frac{e^{\epsilon\gamma_E}}{\Gamma(1-\epsilon)}
		\frac{1}{\epsilon^3}
		\Big[
			2(C_A+2\epsilon\beta_0)\beta_0
		\Big]\nonumber\\
		&& {} - \left(\frac{\mu^2}{-s-i\delta}\right)^{2\epsilon}
		\frac{e^{\epsilon\gamma_E}}{\Gamma(1-\epsilon)}
		\bigg\{
			\frac{1}{\epsilon^3}
			\Big[
				\frac{1}{2}(C_A+4\epsilon\beta_0)\beta_0
			\Big]\nonumber\\
			&& \hspace{2cm}
			- \frac{1}{\epsilon^2}
			\Big[
				\frac{(3\pi^2-67)C_A + 10n_l}{72}(C_A+4\epsilon\beta_0)
			\Big]
			- \frac{1}{\epsilon}
			\Big[
				\frac{1}{2}H_g
			\Big]
		\bigg\}\,,
\end{eqnarray}
with
\begin{eqnarray}
\label{eq::b0Hgdef}
  \beta_0 &=& \frac{1}{4}\left(\frac{11}{3}C_A - \frac{4}{3} T n_l \right)\,,
              \nonumber\\
  H_g &=&
  C_A^2\left(\frac{\zeta_3}{2}+\frac{5}{12}+\frac{11\pi^2}{144}\right)
  + C_A n_l \left( \frac{29}{27} + \frac{\pi^2}{72} \right)
  + \frac{1}{2} C_F n_l + \frac{5}{27} n_l^2
  \,.
\end{eqnarray}



\section{\label{sec::res}Results}

In the following we discuss the results for the finite form factors
constructed according to the prescription of the previous section.  Note that
the one-loop form factors have no dependence on the renormalization scale
$\mu$. At two and three loops the coefficients of the $\log(\mu)$ terms depend
on the choice of the IR subtraction terms.  In our case it is convenient to
cast the results for the two- and three-loop form factors in the following
form
\begin{eqnarray}
  F_X^{{\rm fin},(1)} &=& \tilde{F}_X^{(1)}
              + l_{\mu s} \beta_0 \tilde{F}_X^{(0)}
              \,,\nonumber\\
  F_X^{{\rm fin},(2)} &=& \tilde{F}_X^{(2)}
              + l_{\mu s}
              \left( \beta_1 \tilde{F}_X^{(0)} + 2\beta_0 \tilde{F}_X^{(1)} \right)
              + \beta_0^2 l_{\mu s}^2 \tilde{F}_X^{(0)}
              \,,
\end{eqnarray}
where $\tilde{F}^{(i)}_X = F^{{\rm fin},(i)}_X(\mu^2=-s)$ with $\beta_0$ as defined
in Eq.~(\ref{eq::b0Hgdef}) and
\begin{eqnarray}
  \beta_1 &=& \frac{1}{16}\left(\frac{34}{3} C_A^2 - \frac{20}{3} C_A T n_l  -
              4 C_F T n_l \right)\,,
              \nonumber\\
  l_{\mu s} &=& \log\left(\frac{\mu^2}{-s-i\delta}\right)\,.
\end{eqnarray}
The one- and two-loop results are expanded up to order $1/m_t^{14}$,
the three-loop expressions up to $1/m_t^{8}$ (box) and $1/m_t^{14}$ (triangle).

For illustration we show the analytic result for the leading term ($m_t^0$) of
the three-loop for factors. The corresponding one- and two-loop results
can be found in Ref.~\cite{Davies:2018qvx} and the triangle form factor
up to $1/\mt^{12}$ with numerical values for the colour factors can be found
in Ref.~\cite{Davies:2019nhm}. Our results read
\begin{eqnarray}
\tilde{F}^{(2)}_{\rm tri}&=& {}  \cftwo\Bigg[
\frac{9}{8}
\Bigg]
 + \cfca\Bigg[
\frac{11 \LLmttwos}{12}-\frac{25}{9}
\Bigg]
\nonumber\\&&\mbox{}
 + \catwo\Bigg[
-\frac{7 \LLmttwos}{12}-\frac{253 \zeta (3)}{216}+\frac{\pi ^4}{864}+\frac{19 \pi ^2}{108}+\frac{19777}{3888}
\Bigg]
\nonumber\\&&\mbox{}
 + \cfnl\Bigg[
-\frac{\LLmttwos}{3}+\frac{2 \zeta (3)}{3}-\frac{41}{36}
\Bigg]
 + \canl\Bigg[
-\frac{49 \zeta (3)}{108}-\frac{2255}{1944}-\frac{47 \pi ^2}{1296}
\Bigg]
\nonumber\\&&\mbox{}
 + \cfnh\Bigg[
-\frac{1}{18}
\Bigg]
 + \canh\Bigg[
-\frac{5}{144}
\Bigg]
 + \nltwo\Bigg[
\frac{\pi ^2}{648}
\Bigg]
\,,
\\
\tilde{F}^{(2)}_{\rm box1}&=& - \tilde{F}^{(2)}_{\rm tri} + \cfca\Bigg[
\frac{11}{6}
\Bigg]
 + \catwo\Bigg[
-\frac{7}{6}
\Bigg]
 + \cfnl\Bigg[
-\frac{2}{3}
\Bigg]
 + \cfnh\Bigg[
-1
\Bigg]
\nonumber\\&&\mbox{}
 + \canh\Bigg[
-\frac{2 \mHfour \Litwo\left(1-\frac{\mHfour}{t u}\right)}{9 s^2}-\frac{4
              \mHfour \Litwo\left(\frac{\mHtwo}{t}\right)}{9 s^2}-\frac{4
              \mHfour \Litwo\left(\frac{\mHtwo}{u}\right)}{9 s^2}
\nonumber\\&&\mbox{}
-\frac{1}{9} \Litwo\left(1-\frac{\mHfour}{t u}\right)-\frac{2}{9}
              \Litwo\left(\frac{\mHtwo}{t}\right)-\frac{2}{9}
              \Litwo\left(\frac{\mHtwo}{u}\right)
              -\frac{4 \LLmHtwot \mHfour \log \left(1-\frac{\mHtwo}{t}\right)}{9 s^2}
\nonumber\\&&\mbox{}
-\frac{2}{9} \LLmHtwot \log \left(1-\frac{\mHtwo}{t}\right)-\frac{4 \LLmHtwou
              \mHfour \log \left(1-\frac{\mHtwo}{u}\right)}{9 s^2}-\frac{2}{9}
              \LLmHtwou \log \left(1-\frac{\mHtwo}{u}\right)
\nonumber\\&&\mbox{}
+\frac{11 \LLst}{54}+\frac{11 \LLsu}{54}+\frac{\mHfour \log
              ^2\left(\frac{t}{u}\right)}{9 s^2}+\frac{4 \pi ^2 \mHfour}{27
              s^2}+\frac{2 \mHtwo}{9 s}+\frac{1}{18} \log
              ^2\left(\frac{t}{u}\right)+\frac{2 \pi ^2}{27}
\nonumber\\&&\mbox{}
+\frac{193}{81}
\Bigg]
 + \nlnh\Bigg[
-\frac{\LLst}{27}-\frac{\LLsu}{27}-\frac{10}{81}
\Bigg]\,,
\end{eqnarray}
where $T=1/2$ has been chosen and the overall factor $n_h$ is
contained in Eq.~(\ref{eq::X0}). Furthermore, we have introduced
\begin{eqnarray}
\LLst   &=& \log\left(-\frac{s}{t}\right) - i\pi\,, \nonumber\\
\LLsu   &=& \log\left(-\frac{s}{u}\right) - i\pi\,, \nonumber\\
\LLmttwos &=& \log\left(\frac{m_t^2}{s}\right) + i\pi\,, \nonumber\\
\LLmHtwot &=& \log\left(-\frac{m_H^2}{t}\right) - i\pi\,, \nonumber\\
\LLmHtwou &=& \log\left(-\frac{m_H^2}{u}\right) - i\pi\,.
\end{eqnarray}
We refrain from showing explicit results for $\tilde{F}^{(2)}_{\rm box2}$ which has a
similar structure to $\tilde{F}^{(2)}_{\rm box1}$.  Note that for most colour
structures $\tilde{F}^{(2)}_{\rm box2}$ starts at order $1/m_t^2$ except for the
three colour structures $C_Fn_h$, $C_An_h$ and $n_ln_h$ which arise from
(1PR and 1PI) diagrams with two closed top quark loops.  The analytic results
for the one- and two-loop box and triangle form factors expanded up to $1/m_t^{12}$
and $1/m_t^{14}$ respectively, the
three-loop box form factors $\tilde{F}^{(2)}_{\rm box1}$ and $\tilde{F}^{(2)}_{\rm box2}$
expanded up to $1/m_t^8$, and the three-loop triangle
form factor $\tilde{F}^{(2)}_{\rm tri}$ expanded up to up to $1/m_t^{14}$ can be
found in the supplementary file of this paper~\cite{progdata}.

Note that at two-loop order the 1PI (colour structures $C_ATn_h$ and
$C_FTn_h$) and 1PR ($(Tn_h)^2$) contributions are separately finite. At
three-loop order this is not the case and the whole contribution has to be
considered in order to arrive at finite form factors, see also discussion in
Refs.~\cite{Gerlach:2018hen,Grigo:2014jma}.

Let us now briefly discuss the numerical impact of our
calculation.  For the numerical evaluation we use $m_t=173$~GeV and
$m_H=125$~GeV and for the transverse momentum we introduce the parameter
\begin{eqnarray}
  r_{p_T} &=& \frac{p_T^2}{s}\,,
\end{eqnarray}
with $r_{p_T}=0.01$ as default value.
Furthermore we choose for the parameters introduced for closed fermion loops
$n_l=5$ and $n_h=1$.

\begin{figure}[t]
  \hspace*{-1em}
  \begin{tabular}{cc}
    \includegraphics[width=0.49\linewidth]{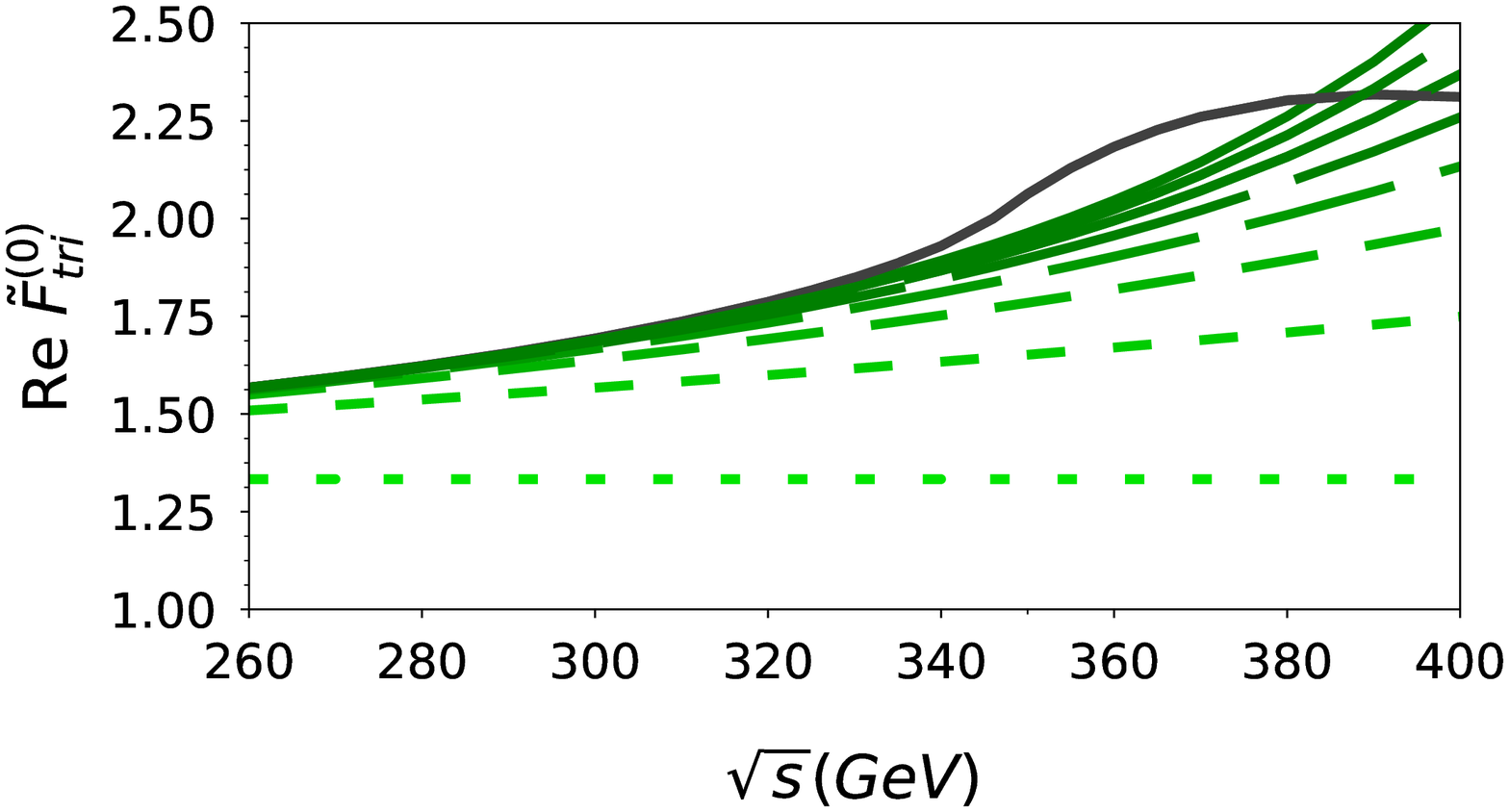} &
    \includegraphics[width=0.49\linewidth]{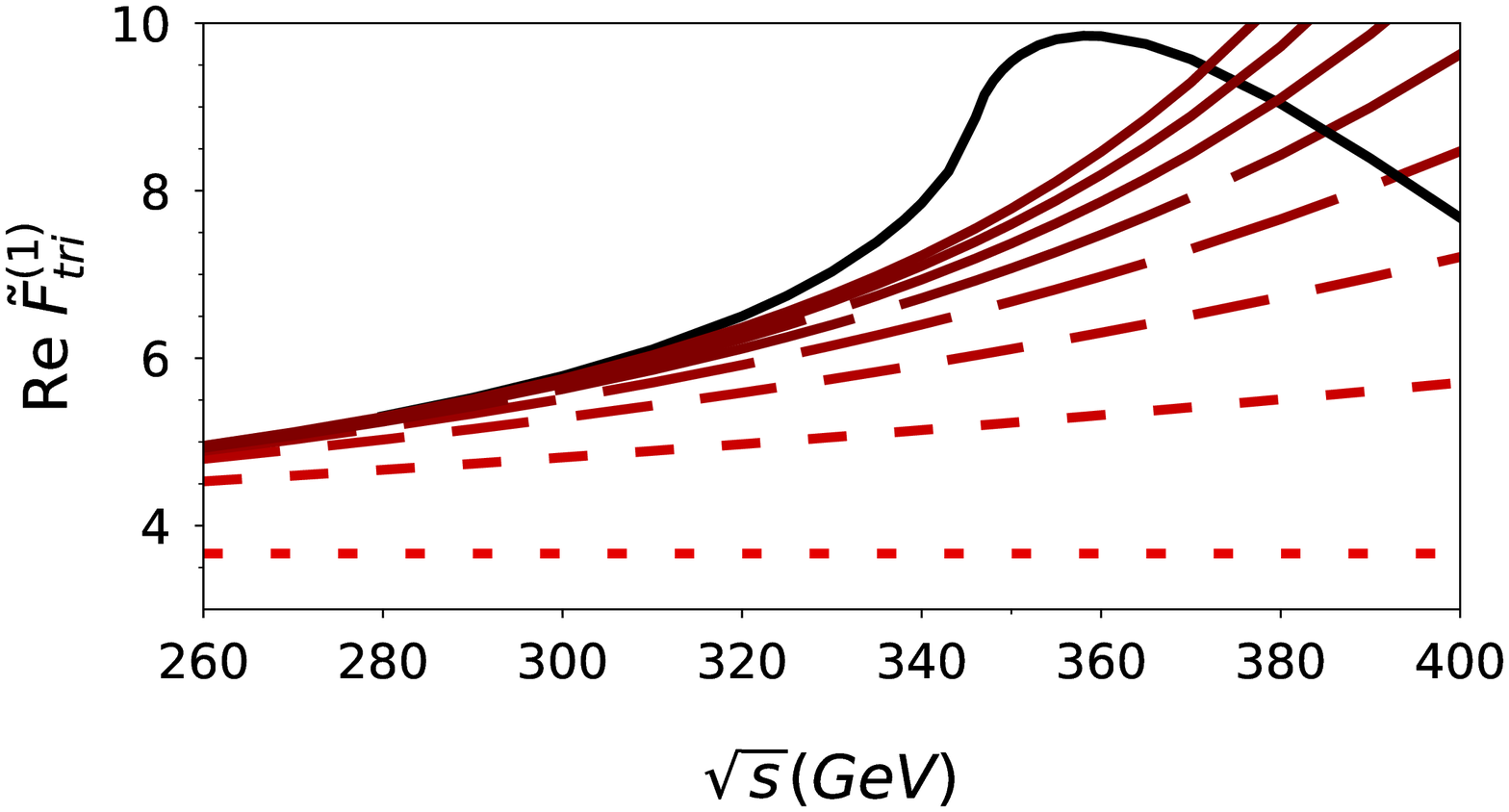}
    \\
    \includegraphics[width=0.49\linewidth]{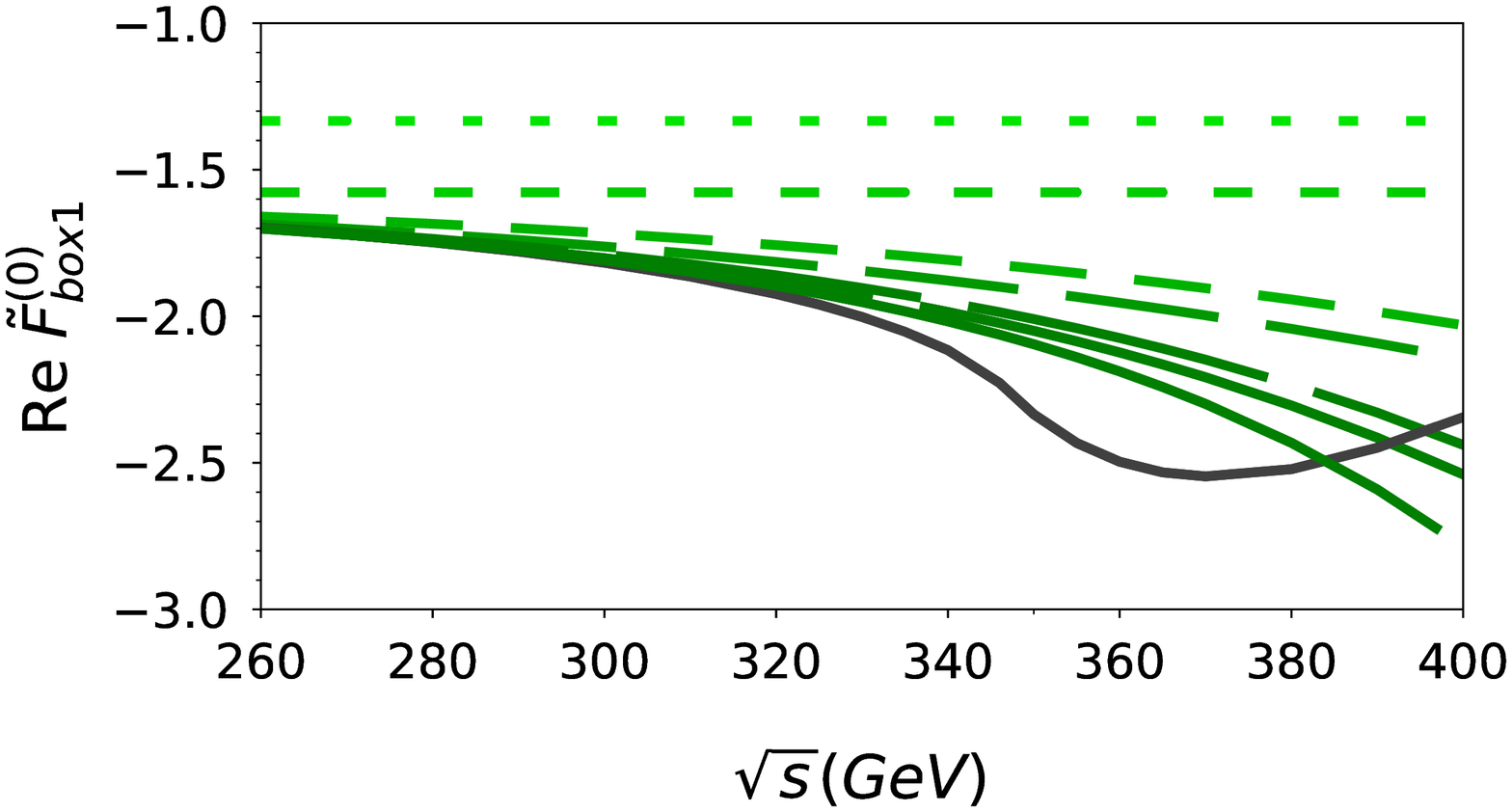} &
    \includegraphics[width=0.49\linewidth]{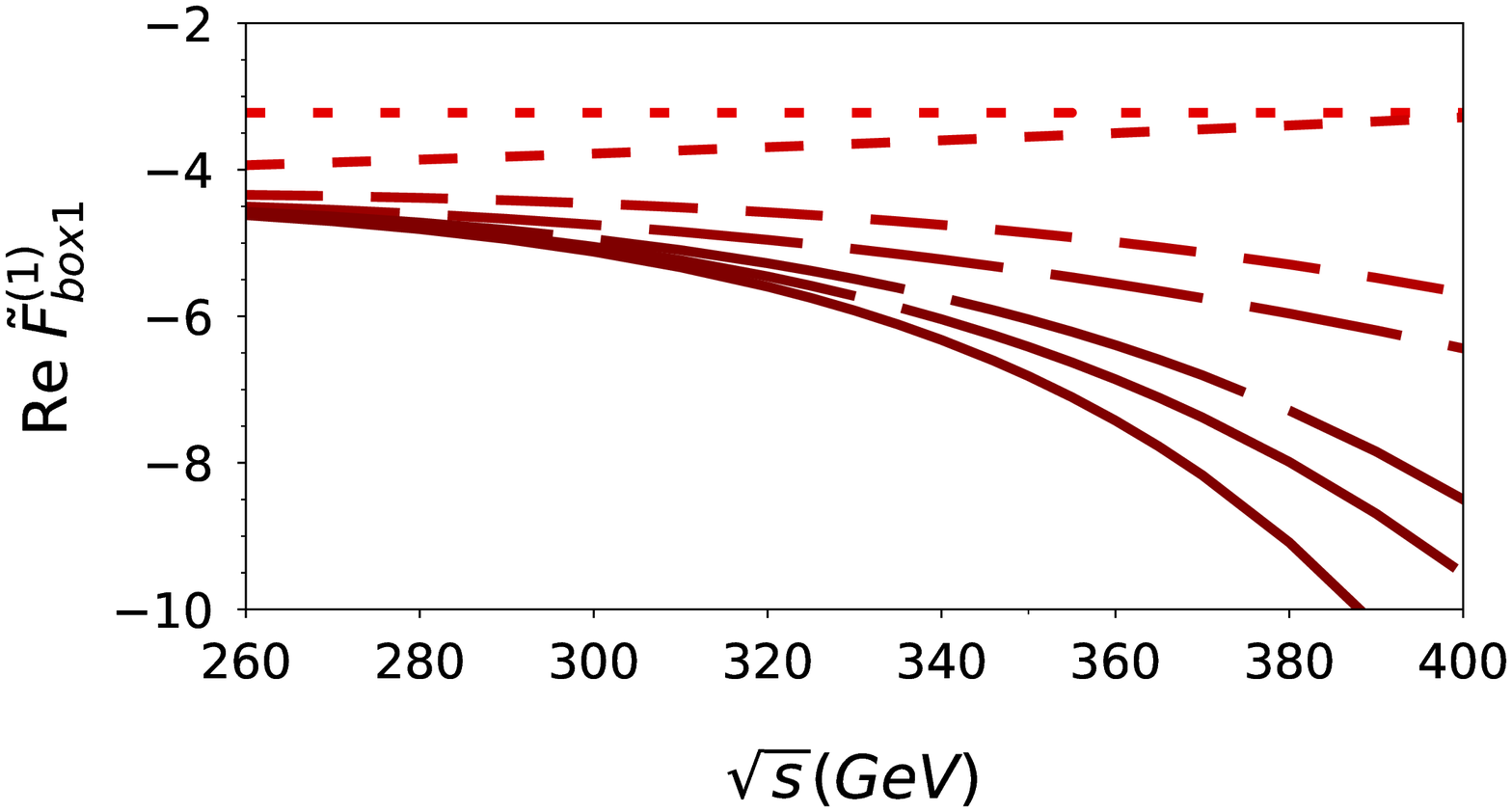}
    \\
    \includegraphics[width=0.49\linewidth]{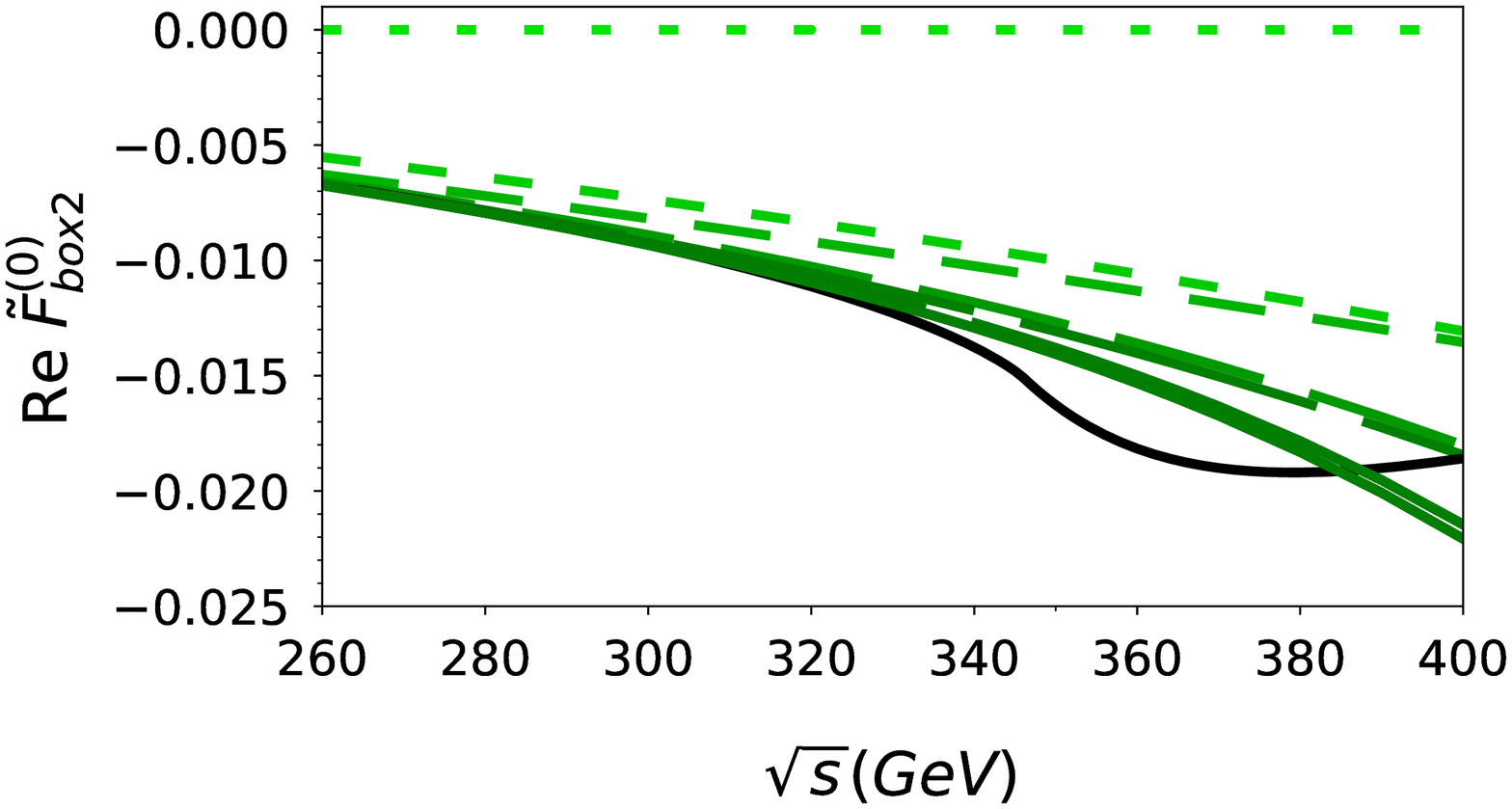} &
    \includegraphics[width=0.49\linewidth]{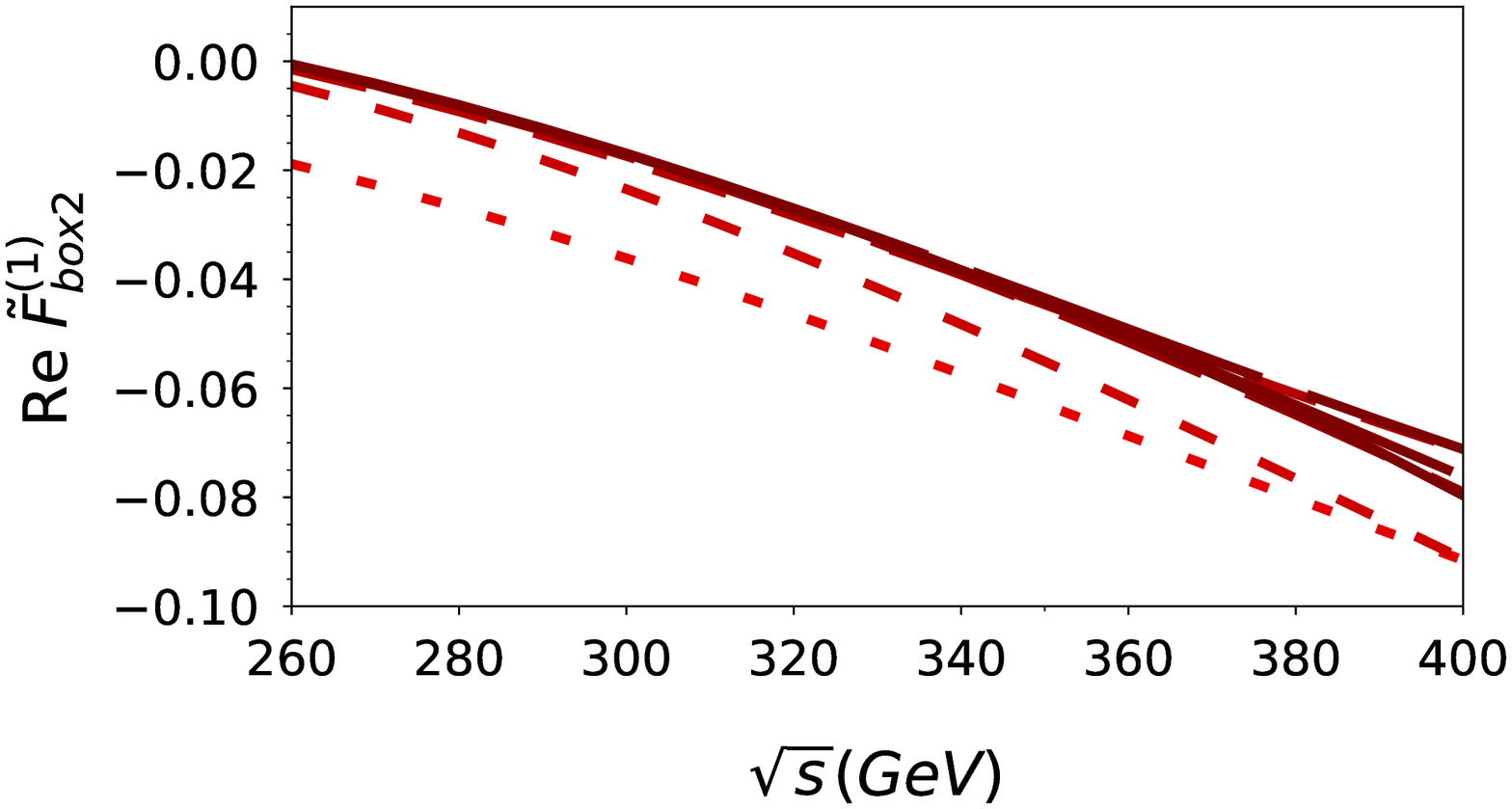}
  \end{tabular}
  \caption[]{\label{fig::ff}Real parts of one- and two-loop form factors as a
    function of $\sqrt{s}$ for $r_{p_T}=0.01$.}
\end{figure}

In Fig.~\ref{fig::ff} we show real parts of the one- and two-loop results for
the three form factors as a function of $\sqrt{s}$. We include terms up to
order $1/m_t^{14}$ for the triangle and order $1/m_t^{12}$ for the box
form factors. Lines with longer dashes include more expansion terms.  Below
the threshold, i.e. for $\sqrt{s}\lsim 2m_t$, one observes a reasonable
convergence of the expansion in $1/\mt^2$ as can be seen by the reduced
distance between the dashed curves. In this respect, $F^{(0)}_{\rm box2}$ and
$F^{(1)}_{\rm box2}$ are particularly well-behaved; after including the third
expansion term the curves lie practically on top of each other. At one-loop
order we also find good agreement with the exact results (solid black curves)
for $\sqrt{s}\lsim 320$~GeV.  For the two-loop triangle form factor we find
agreement with the exact expression (see
Refs.~\cite{Harlander:2005rq,Anastasiou:2006hc,Aglietti:2006tp} for analytic
expressions) for $\sqrt{s}\lsim 300$~GeV.

Note that the form factors also develop imaginary parts which 
originate from contributions with massless cuts,
see co-subgraphs in Fig.~\ref{fig::diags_ae}. They are contained in
our analytic expressions~\cite{progdata}, but are not plotted here.

\begin{figure}[t]
  \hspace*{-1em}
  \begin{tabular}{c}
    \includegraphics[width=0.49\linewidth]{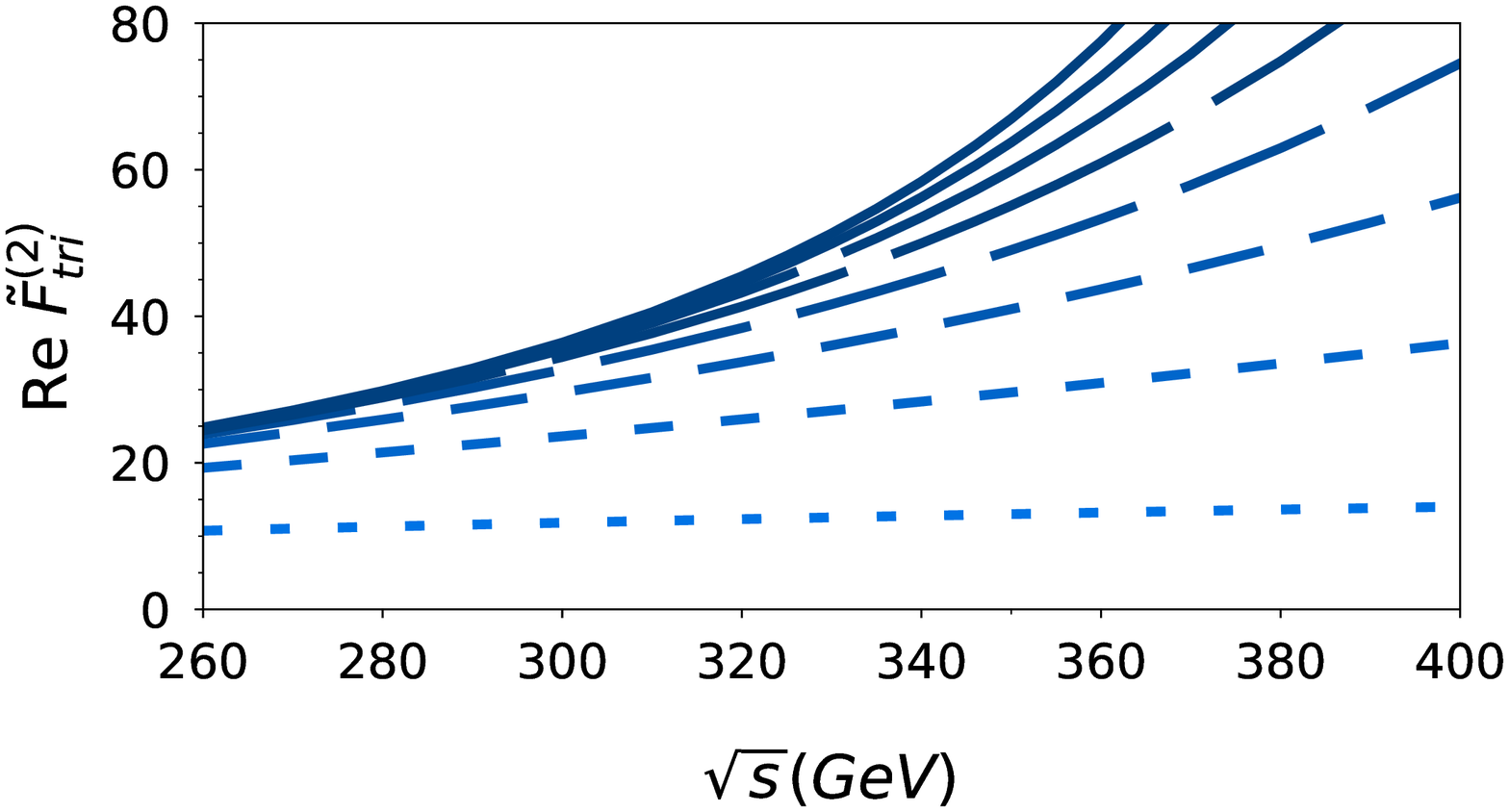}
    \includegraphics[width=0.49\linewidth]{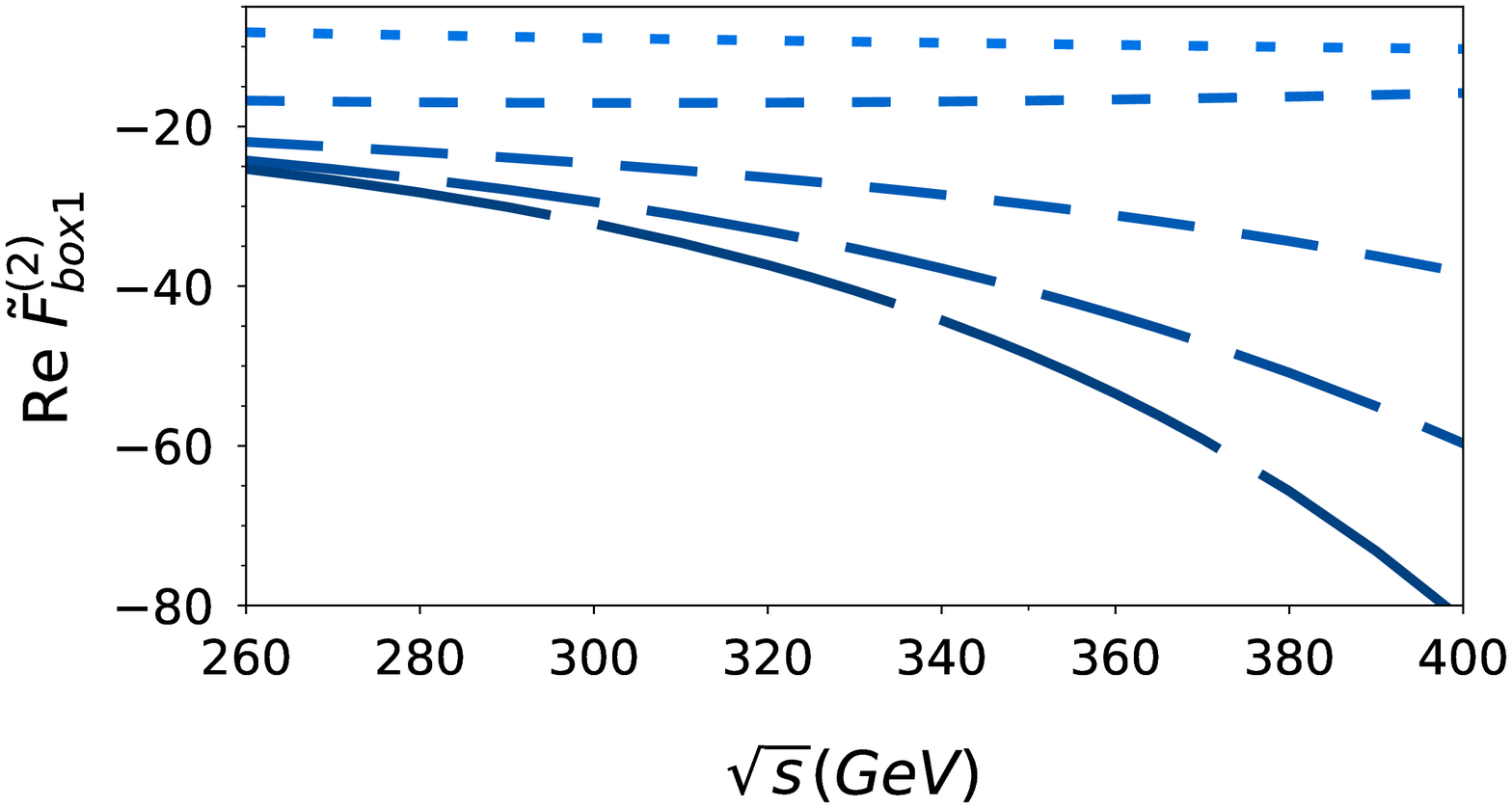}
    \\
    \includegraphics[width=0.49\linewidth]{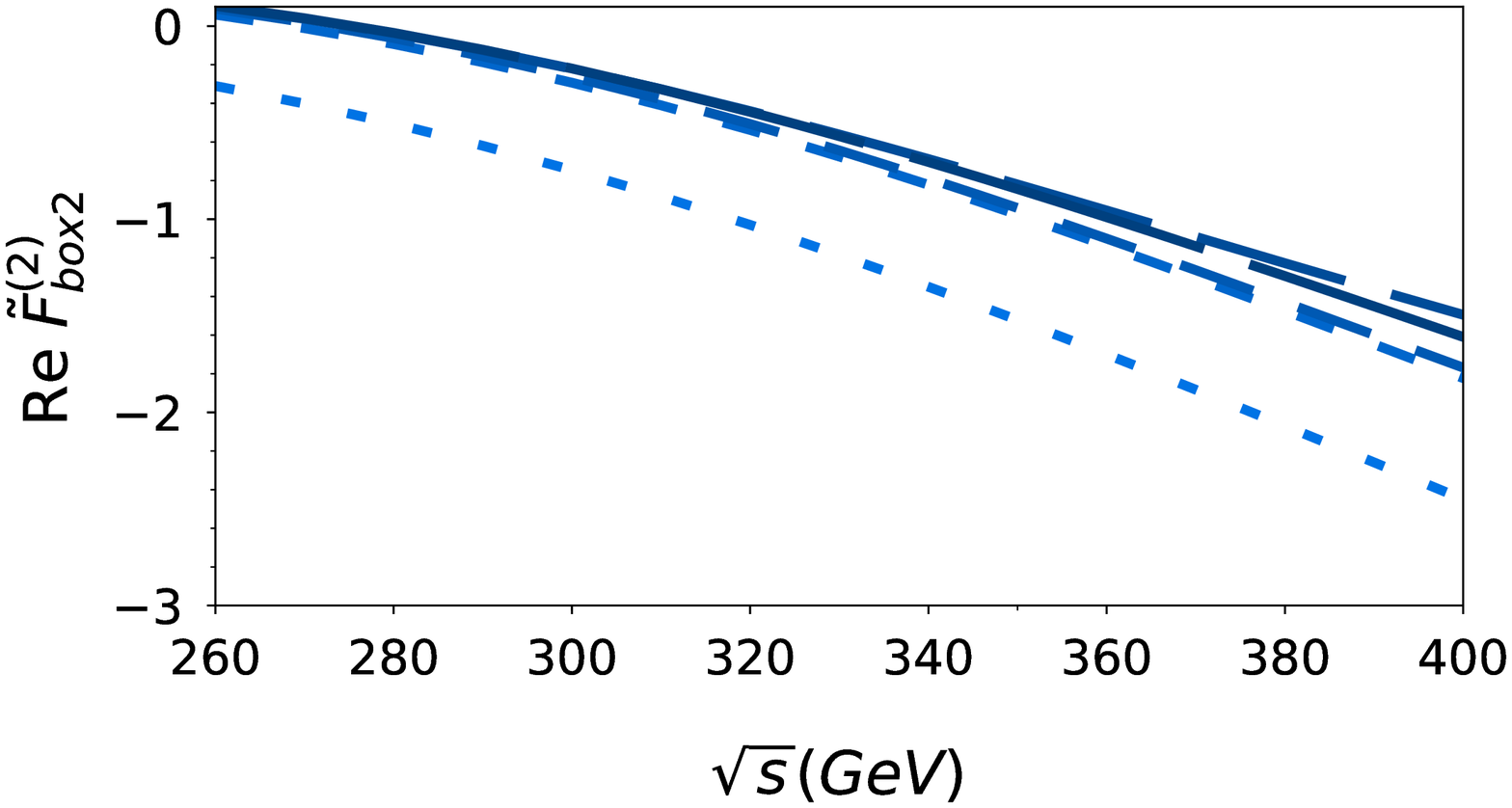}
  \end{tabular}
  \caption[]{\label{fig::ff3l}Real parts of three-loop form factors as a
    function of $\sqrt{s}$ for $r_{p_T}=0.1$.}
\end{figure}

A similar behaviour to the one- and two-loop cases is observed at three-loop
order as can be seen in Fig.~\ref{fig::ff3l}.  We want to stress that
qualitatively the two- and three-loop corrections show a very similar
behaviour. Since the two-loop terms have proven to provide useful and
important input into the Pad\'e procedure~\cite{Grober:2017uho}, which can be
used to obtain approximations valid in the whole $\sqrt{s}$ region, we expect
that the three-loop terms are of similar importance.

For some applications it is advantageous to rescale the
higher order corrections by the exact leading order contributions
using
\begin{eqnarray}
  \frac{ \tilde{F}_X^{(n),\rm exp} }{ \tilde{F}_X^{(0),\rm exp} } \tilde{F}_X^{(0),\rm exact} 
  \,,
  \label{eq::rescale}
\end{eqnarray}
where $\tilde{F}_X^{(n),\rm exp}$ and $\tilde{F}_X^{(0),\rm exp}$ are expanded
up to the same order in $1/m_t$. We refrain from showing the corresponding results
but simply want to mention that the differences  between the 
$\tilde{F}_X^{(n),\rm exp} $ and the rescaled expression~(\ref{eq::rescale})
become smaller with increasing order in $1/m_t$.
In fact the curves which correspond to the deepest expansions are
very close to each other.


\section{\label{sec::con}Conclusions}

We compute three-loop corrections to the process $gg\to HH$ in the large-$m_t$
limit and provide results for five expansion terms (up to order $1/m_t^8$) for
the two box-type form factors and for eight expansion terms (up to order
$1/m_t^{14}$) for the triangle form factor. As compared to previous
work~\cite{Grigo:2015dia} we have computed two\footnote{three for the
  triangle form factor} more expansion terms, which required significant
reorganization and optimization of the calculations since huge expressions are
obtained at various intermediate stages.  We discuss these modifications
in Section~\ref{sec::calc}.
Furthermore in Ref.~\cite{Grigo:2015dia} only partonic cross sections, rather than
individual form factors, are available.

The analytic results for the form factors, which are provided in a computer-readable
form in supplementary material~\cite{progdata}, are useful input for the construction
of approximations for $gg\to HH$ at NNLO, both for total cross sections and differential
distributions.  This concerns both the construction of Pad\'e approximants along the lines
of~\cite{Grober:2017uho} (indeed these new results for the triangle form factor have already
been used in~\cite{Davies:2019nhm}) but also approximation procedures which have been
employed in Ref.~\cite{Grazzini:2018bsd}.

 
\section*{Acknowledgements}

We thank Ramona Gr\"ober, Florian Herren, Andreas Maier, Go Mishima, Thomas
Rauh and David Wellman for fruitful discussions and comments.  This work is supported by
the Collaborative Reseach Center 257 ``P$^3$H: Particle Physics Phenomenology
after the Higgs Discovery''.


\begin{appendix}


\section{README for the supplementary material}

In this Appendix we provide a brief explanation of the
notation used in the ancillary file to this paper~\cite{progdata}.

Our final results for the form factors are contained in the file {\tt
  resFF.m} where the following notation has been used:
\\[-.5em]
\begin{center}
\begin{tabular}{ccc|ccc|ccc}
  {\tt F1tri}  & {\tt F2tri}   & {\tt F3tri} & 
  {\tt F1box1} & {\tt F2box1}  & {\tt F3box1} &
  {\tt F1box2} & {\tt F2box2}  & {\tt F3box2}
  \\
  \hline
  $\tilde F^{(0)}_{\rm tri}$ & $\tilde F^{(1)}_{\rm tri}$ & $\tilde F^{(2)}_{\rm tri}$ &
  $\tilde F^{(0)}_{\rm box1}$ & $\tilde F^{(1)}_{\rm box1}$ & $\tilde F^{(2)}_{\rm box1}$ &
  $\tilde F^{(0)}_{\rm box2}$ & $\tilde F^{(1)}_{\rm box2}$ & $\tilde F^{(2)}_{\rm box2}$
  \\
\end{tabular}
\end{center}

The expressions have the same colour factors as in Eq.~(\ref{eq::colfac})
where $T=1/2$ has been chosen and the overall factor $n_h$ has been set to 1.
The following variables are used
\verb|ca, cf, nh, nl, mH2| $=m_H^2$ \verb|mt2|$=m_t^2$, \verb|s, t|.
Furthermore the functions \verb|Li2[_]| and \verb|Log[_]|
are used with obvious meaning.


\end{appendix}

\end{document}